\documentclass[preprintnumbers,superscriptaddress,amsmath,amssymb,10pt,twocolumn,a4paper]{revtex4-1}

\usepackage{geometry}
\usepackage{ulem}
\usepackage{mathtools}
\usepackage{graphicx}
\usepackage{dcolumn}
\usepackage{bm}
\usepackage{makecell}
\usepackage{xcolor}
\usepackage{enumerate}
\usepackage[caption=false]{subfig}
\usepackage{float}
\usepackage{soul}
\usepackage{array}
\usepackage[unicode=true,pdfusetitle,
 bookmarks=true,bookmarksnumbered=false,bookmarksopen=false,
 breaklinks=false,pdfborder={0 0 1},backref=false,colorlinks=true]
 {hyperref}
\hypersetup{linkcolor=blue,urlcolor=blue,citecolor=blue}

\newcommand{\be}{\begin{equation}}
\newcommand{\ee}{\end{equation}}
\newcommand{\bd}{\begin{displaymath}}
\newcommand{\ed}{\end{displaymath}}
\newcommand{\BE}{\begin{eqnarray}}
\newcommand{\EE}{\end{eqnarray}}

\newcommand{\bx}{\ensuremath{\mathbf{x}}}

\newcommand{\bn}{\ensuremath{\mathbf{n}}}

\newcommand{\avg}[1]{\left\langle{#1}\right\rangle}

\begin{document}

\title{Ordering dynamics of nonlinear voter models}

 \author{Luc\'ia S. Ramirez}
\email{luciaramirez@ifisc.uib-csic.es}
\affiliation{Instituto de F\'isica Interdisciplinar y Sistemas Complejos, IFISC (CSIC-UIB), Campus Universitat Illes Balears, E-07122 Palma de Mallorca, Spain}

\author{Federico Vazquez}
\email{fede.vazmin@gmail.com}
\affiliation{Instituto de C\'alculo, FCEyN, Universidad de Buenos Aires and Conicet, Intendente Guiraldes 2160, Cero + Infinito, Buenos Aires C1428EGA, Argentina}

\author{Maxi San Miguel}
\email{maxi@ifisc.uib-csic.es}
 \affiliation{Instituto de F\'isica Interdisciplinar y Sistemas Complejos, IFISC (CSIC-UIB), Campus Universitat Illes Balears, E-07122 Palma de Mallorca, Spain}
 
 \author{Tobias Galla}
\email{tobias.galla@ifisc.uib-csic.es}
\affiliation{Instituto de F\'isica Interdisciplinar y Sistemas Complejos, IFISC (CSIC-UIB), Campus Universitat Illes Balears, E-07122 Palma de Mallorca, Spain}

\date{\today}

\begin{abstract} 
 We study the ordering dynamics of nonlinear voter models with multiple states, also providing a discussion of the two-state model. The rate with which an individual adopts an opinion scales as the $q$-th power of the number of the individual's neighbours in that state.  For $q>1$ the dynamics favor the opinion held by the most agents. The ordering to consensus is driven by deterministic drift, and noise only plays a minor role. For $q<1$ the dynamics favors minority opinions, and for multistate models the ordering proceeds through a noise-driven succession of metastable states. Unlike linear multi-state systems, the nonlinear model cannot be reduced to an effective two-state model. We find that the average density of active interfaces in the model with multiple opinion states does not show a single exponential decay in time for $q<1$, again at variance with the linear model. This highlights the special character of the conventional (linear) voter model, in which deterministic drift is absent. As part of our analysis, we develop a pair approximation for the multi-state model on graphs, valid for any positive real value of $q$, improving on previous approximations for nonlinear two-state voter models.\\

\end{abstract}
\maketitle


\section{Introduction}
 The voter model (VM) is a popular mathematical model, inspired by opinion dynamics and social interactions~\cite{clifford,liggett1,holey,castellano_fortunato,redner,suchecki}. One focus of interest is if consensus emerges from the imitation dynamics of the VM, or if diversity prevails.

The VM describes a set of agents represented as nodes in a contact network. In the simplest setting, each agent is characterized by a binary opinion state. At each time step, an agent is selected at random, and adopts the opinion of one of its neighbors. The neighbor is also selected randomly. 
With these dynamics, the model has two absorbing states. In these states consensus has been reached, that is, all agents hold the same opinion. Therefore, no more opinion changes are possible through imitation.

The simple two-state VM is a starting point for more complex opinion dynamics models. Its simplicity allows for multiple variations that capture different features of social interaction \cite{castellano_fortunato, redner,suchecki}. One of these is the inclusion of `noise', that is agents can spontaneously change their opinion~\cite{granovsky,carro2016,peralta}. Noisy VMs are closely related to the Kirman model of herding in financial
markets~\cite{kirman}. Further extensions of the basic VM include models with more than two opinion states  \cite{starnini,msvm, dall, herrerias,sznajd_qmsvm,vazquez-2019,vazquez-2003,vazquez-2004,mobilia2011, castello2006,loscar-2021}. In some of these models the opinion states are equivalent to one another in the sense that no opinion has any preferential status over another~\cite{starnini,herrerias,vazquez-2019,sznajd_qmsvm,loscar-2021,msvm}. In other multi-state models the different options are not equivalent~\cite{vazquez-2003,vazquez-2004,dall,mobilia2011, castello2006}. The basic VM can also be extended to include nonlinear interactions ~\cite{castellano_qvm,schweitzer2009,sznajd_qvm,peralta2018, timapanaro2014,  mobilia2015, mobilia2016,jed_qvm, sznajd_qvm, min, min2, raducha}.

It is of particular interest to physicists how these different features affect the ordering dynamics, and in particular quantities such as the time to reach consensus in finite systems \cite{castellano_fortunato, dornic, redner,sood-2005,sood,ben,suchecki,vazquez_eguiluz}. A typical order parameter for the study of ordering dynamics and the path to consensus is the density of `active links'. A link in the interaction network is said to be `active' if it connects agents who are in different opinion states. In the binary VM, the mean density of active links can be shown to have an exponential decay in time, both for models with complete interaction graphs \cite{ben}, and on uncorrelated graphs  \cite{vazquez_eguiluz}. This exponential decay has a characteristic time, $\tau$, that depends on the system size $N$ and on the graph. For instance,  $\tau \sim N$ for the complete graph (CG), and Erd\"os–Renyi networks (ER) \cite{cast, vazquez_eguiluz}, and $\tau \sim  N / \ln N$ for Barab\'asi--Albert (BA) graphs \cite{sood-2005, vazquez_eguiluz}. 
Recently, some of the current authors showed that the density of active links decays exponentially also in multistate VMs \cite{msvm}. The decay times were found to be the same as in the binary VM.

In this paper, we focus on nonlinear voter models (NLVM). The word nonlinear is here to be interpreted as follows: In the standard (linear) VM the probability with which an agent changes to a particular opinion state is proportional to the fraction of its neighbours in that state.  In the NLVMs this probability is instead a nonlinear function of the fraction of neighbours in a particular state \cite{castellano_qvm,schweitzer2009}. Specifically, we consider settings in which the probability of switching to a particular opinion state is proportional to the $q$-th power of the fraction of neighbors in that state. For $q=1$, this reduces to the linear VM.

In much of the recent work on the NLVM  \cite{timapanaro2014,  mobilia2015, mobilia2016,jed_qvm, sznajd_qvm, sznajd_qmsvm}
$q$ takes positive integer values. In this case the dynamics can be interpreted as follows: A node consults with a panel of $q$ agents, selected at random from its neighbors (with replacement). If all $q$ neighbors agree with one another and are in a state that is different to that of the focal agent, the focal agent takes on that opinion. However, more generally, one can consider $q$ as a mathematical parameter taking any positive real value \cite{vazquez2010,peralta2018,min,raducha,min2} as we do in this paper. In this second wider interpretation, also used in social impact theory \cite{nowak}, $q$ measures the nonlinear effect of local majorities which is qualitatively different for $q<1$ or $q>1$. The distinction between these two regimes is one key aspect of the present work.

The main objective of this paper is to characterise the dynamics of NLVM with multiple opinion states, and for general values of the parameter $q$. While the linear multistate VM ($q = 1$) can be reduced to an effective  binary VM for many purposes \cite{redner, herrerias, msvm}, this is not the case for the multi-state NLVM (as we will show below). Consequently, one can expect multi-state NLVMs to behave quite differently than their two-state counterparts. We will explore these differences both at the level of individual realisations and as an ensemble average.  

The remainder of the paper is organized as follows.  In Sec.~\ref{sec:model} we define the multi-state NLVM model, and we introduce the main macroscopic quantities of interest. We also highlight some of the behavior of the models $q<1$, $q=1$, and $q>1$ respectively. Section~\ref{sec:2_states_all-to-all} focuses on the dynamics on complete interaction graphs for the case of two opinion states. We here summarise existing results, but also present additional perspectives.  The focus of Sec.~\ref{sec:multi_state} is on multistate model with all-to-all interaction. We study both individual realisations and the ensemble average, characterising the time-evolution of different quantities by means of numerical simulations and a mean-field approach.  Sec.~\ref{sec:networks} deals with the dynamics of the binary model on complex networks by means of the pair approximation. In Sec.~\ref{sec:pa_qmsvm} we then  investigate the multi-state model on complex networks, and develop an analytical description within the pair approximation. Finally, Sec.~\ref{sec:conclusions} contains a summary, and our conclusions. Further details of our analysis can be found in the appendix.

\section{Model definitions}\label{sec:model}

\subsection{Setup and notation}
The model describes a population of $N$ individuals who can each be in one of $M$ discrete states (the `opinions'). Individuals are labeled $i=1,\dots,N$, and states $\alpha=1,\dots,M$. The variable $s_i(t)\in\{1,\dots,M\}$ represents the state of individual $i$ at time $t$.

The individuals are located at the nodes of a static undirected interaction network, and each individual can interact with its nearest neighbours. We use the notation $c_{ij}$ for the adjacency matrix, with $c_{ij}=c_{ji}=1$ if individuals $i$ and $j$ are neighbors, and $c_{ij}=c_{ji}=0$ if they are not. We also set $c_{ii}=0$. We write $j\in i$ to indicate that $j$ is among the neighbors of $i$. The degree of  individual $i$ is denoted by $k_i$ ($k_i=\sum_j c_{ij}$). The total number of links in the graph is $E=\sum_{i<j} c_{ij}$.  In a complete graph (CG), i.e., a model with all-to-all interactions ($c_{ij}=1$ for all $i\neq j$), we have $E=N(N-1)/2$.

We denote the number of individuals holding opinion $\alpha$ by $n_\alpha$, and we write $\bn(t)=[n_1(t),\dots, n_M(t)]$. We also introduce $x_\alpha = n_\alpha/N$ as the fraction of individuals in opinion state $\alpha$.

Throughout the paper we use the notation $\avg{\dots}$ to indicate averages over realisations of the dynamics.

\begin{figure*}[t]
\centering
\includegraphics[width=0.9\textwidth]{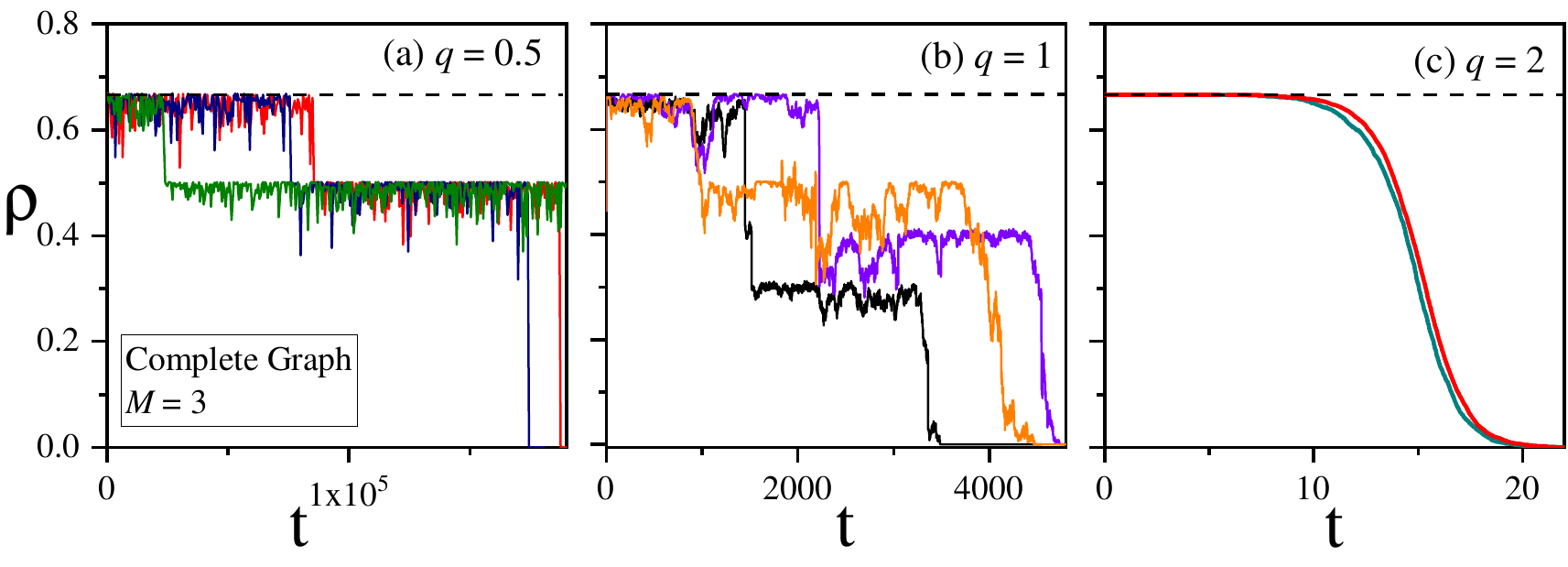}
\caption{Density of active links in individual realizations of the multistate NLVM on complete graphs, and with initially $3$ opinions ($M = 3$). Panel (a): $q=0.5$ and $N=1000$, (b)  $q=1$ (linear VM) and $N=4000$, and (c)  $q=2$ and $N=10000$. Each of the colored curves in the different panels represents one realisations. The dotted line indicates the initial density of active links ($\rho=2/3$) for balanced initial conditions.\label{fig:1realisations_cg}}
\end{figure*}

\subsection{Dynamics and transition rates} \label{sec:dynamics}
In the most common set-up of the VM, individuals change their opinion through an imitation process. The dynamics consists of the following two steps: (i) randomly select an individual $i$; then (ii) randomly select a neighbour $j\in i$, and if the two individuals are in different states, $i$ will copy the state of $j$. 

In a continuous-time setting of the basic VM the rate at which individual $i$ adopts opinion state $\alpha$ is given by
\be
\label{eq:rate_lineal}
T_{i,\alpha}=\frac{n_{i,\alpha}}{k_i},
\ee
where 
\be
n_{i\alpha}\equiv\sum_{j\in i} \delta_{s_j,\alpha}
\ee
is the number of neighbors of node $i$ who are in opinion state $\alpha$ ($\delta$ is the Kronecker delta).  The rate for $i$ to change to $\alpha$ is thus linear in $n_{i\alpha}$.

We consider a nonlinear version of these dynamics with group interaction. Instead of copying the state of a randomly selected neighbour, the rate of changing into state $\alpha$ is proportional to the $q$-th power of the density of neighbours in state $\alpha$.

Mathematically (and again assuming time is continuous), the rate with which individual $i$ adopts opinion state $\alpha$ is
\be
\label{eq:rate}
T_{i,\alpha}=\left(\frac{n_{i,\alpha}}{k_i}\right)^q.
\ee
These rates define the multistate nonlinear voter model (multistate NLVM). For $q=1$ this reduces to the conventional (linear) multistate VM.

\subsection{Density of active links}

Each link in the interaction network is either `active' or `inactive' at any given time. A link $(i,j)$ is considered active when the two nodes it connects are in different states ($s_i\neq s_j$); otherwise, the link is inactive. 

We define the fraction of links of type $\alpha\beta$,
\be\label{eq:rhoalphabeta}
\rho_{\alpha\beta}=\frac{1}{E}\sum_{i<j} c_{ij} \left(\delta_{s_i,\alpha}\delta_{s_j,\beta}+\delta_{s_j,\alpha}\delta_{s_i,\beta}\right),
\ee
for $\alpha\neq\beta$. The total density of active links in the system is then
\be\label{eq:defrho}
\rho=\sum_{\alpha<\beta} \rho_{\alpha\beta}=\frac{1}{E}\sum_{i<j} c_{ij} (1-\delta_{s_i,s_j}).
\ee

\subsection{Significance of the parameter \textit{q}}

In our model, the parameter $q$ can take any positive real value. This reflects nonlinear interactions as considered for example in social impact theory \cite{nowak}, and in problems of language competition \cite{vazquez2010,abrams,nettle}. 

Very different dynamics are expected for $q<1$ and $q>1$, respectively. Relative to the linear model, values of $q<1$ favour co-existence in the two-state nonlinear VM, and one finds long-lived metastable states. On the other hand, consensus is promoted for $q>1$. For further details see for example \cite{vazquez2010,min}.

An illustration of the behaviour of multi-state nonlinear VMs can be found in Fig.~\ref{fig:1realisations_cg}. We show how the density of active links evolves in time in different realisations of the model on a complete graph and with initially $M=3$ opinion states. All realisations were initialized with what we will call `balanced' initial conditions, that is to say at $t=0$ a third of the agents hold each of the three opinion states. 

For $q=0.5$ [Fig.~\ref{fig:1realisations_cg}(a)] we find that the density of active links goes through cascades of plateaux for each realization. Initially all realisations start with three opinion states, and remain near $\rho\approx 2/3$. The different realizations reside at this initial plateau for different amounts of time. All realizations eventually transit to an intermediate plateau at $\rho \approx 0.5$, and eventually end at a consensus state. The times at which this intermediate plateau is reached varies across realizations, as does the departure time. In the linear VM $q=1$, [panel (b)] realizations also go through intermediate plateaux, but the value of the density of active links at each plateau differs from realization to realization \cite{msvm}. Fig.~\ref{fig:1realisations_cg}(c) finally is for $q=2$. No plateaux are observed, and the path to consensus appears far less noisy than in panels (a) and (b). We note though that the time of departure from the initial density of active links ($\rho=2/3$) varies across realizations in panel (c). We also highlight that, despite the larger size of the network in panel (c), the time to consensus is much shorter than for $q<1$ [panel~(a)].  

\subsection{Master equation for the model on a complete graph}\label{sec:simplify_CG}

When any node can interact with any other node, the position of an individual on the interaction network becomes irrelevant. The state of the system is fully described by the vector $\bn=(n_1,\dots,n_M)$. 

The rates in Eq.~(\ref{eq:rate}) then reduce to 
\be
T_{i,\alpha} = \left(\frac{n_\alpha}{N-1}\right)^q,
\ee
assuming node $i$ is not already in state $\alpha$. The total rate for conversion of individuals in state $\beta$ to state $\alpha$ in the population is 
\be\label{eq:rates_aa}
{\cal T}_{\beta\to\alpha}(\bn)=n_\beta\left(\frac{ n_\alpha}{N-1}\right)^q.
\ee
The time evolution of the probability $P(\bn)$ of finding the system in state $\bn$ fulfils the master equation
\be\label{eq:master_all_to_all}
\frac{d}{dt}P(\bn)=\sum_{\alpha\neq\beta} (E_\alpha E_\beta^{-1}-1)[{\cal T}_{\alpha\to\beta}(\bn)P(\bn)].
\ee
We have here defined the raising operator $E_\alpha$. It acts on functions of $\bn$ increasing the argument $n_\alpha$ by one, i.e., we have $E_\alpha f(\bn)=f(n_1,\dots, n_\alpha+1,\dots, n_M)$. The inverse operator is written as $E_\alpha^{-1}$.

\section{Two-state NLVM with all-to-all interactions}
\label{sec:2_states_all-to-all} 

In this section we focus on the model with $M=2$ opinion states on complete interaction graphs \cite{vazquez2010,mobilia2015,peralta2018}. We review the setup, the resulting rate equations and their fixed-point structure. 
We also present an estimate of the time-to-consensus for $q<1$, based on an Arrhenius law for the escape from the minimum of a potential. Further, we present a more detailed analytical study of the special cases $q=2$ and $q=3$.


\subsection{Rate equations and different dynamical regimes for $q<1$, $q=1$ and $q>1$}\label{sec:rate_eq}
If there are only two opinion states ($\alpha=1,2$), then the state of the system on a CG is fully specified by the number $n$ of individuals in opinion state $1$. All remaining $N-n$ agents have opinion $2$. Alternatively (and using the analogy with spin systems), the state of the system can be characterised by the magnetisation $m=2(n/N) -1$. 

Events in the population either increase or decrease the state of the system $n$ by one. The rates ${\cal T}_n^\pm$ for these events can be read off from Eq.~(\ref{eq:rates_aa}),
\be \label{eq:rate_cg_m2}
{\cal T}_n^+=(N-n) \left(\frac{n}{N-1}\right)^q, ~~{\cal T}_n^-=n\left(\frac{N-n}{N-1}\right)^q. 
\ee
The states $n=0$ and $n=N$ are absorbing.

In the limit of an infinite population ($N\to\infty$) one finds the rate equation
\be \label{eq:rate_binary}
\dot {x}=(1- x)x^q- x (1- x)^q,
\ee
where $x=n/N$ is the proportion of agents in opinion state $1$. The density of active links in the model with two states on a CG is $\rho=2x(1-x)$.

The fixed points of Eq.~(\ref{eq:rate_binary}) are $x^*=0$, $x^*=1$ and $x^*=1/2$. It is useful to write Eq.~(\ref{eq:rate_binary}) in the form
\be
    \frac{dx}{dt} = - \frac{\partial V(x)}{\partial x}
\label{eq:rate_binary_Potentiasl}
\ee  
with the potential
\begin{eqnarray}
    V(x) &=& \frac{\left[ x^{(q+2)}+(1-x)^{(q+2)}-1 \right]}{(q+2)}  \nonumber \\
    &-& \frac{\left[ x^{(q+1)}+(1-x)^{(q+1)}-1 \right]}{(q+1)}.
    \label{eq:V-x}
\end{eqnarray}
A Landau expansion of this potential is given in \cite{vazquez2010}.  The fixed points correspond to the extrema of $V(x)$.

\begin{figure}[t]
\centering
{\includegraphics[width=0.85\columnwidth]{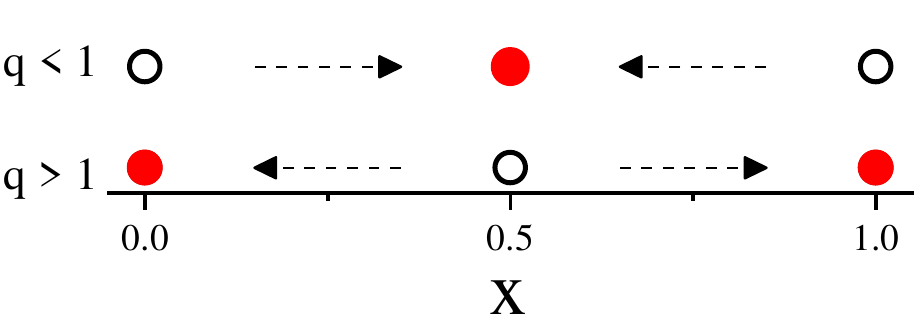}}
{\includegraphics[width=1\columnwidth]{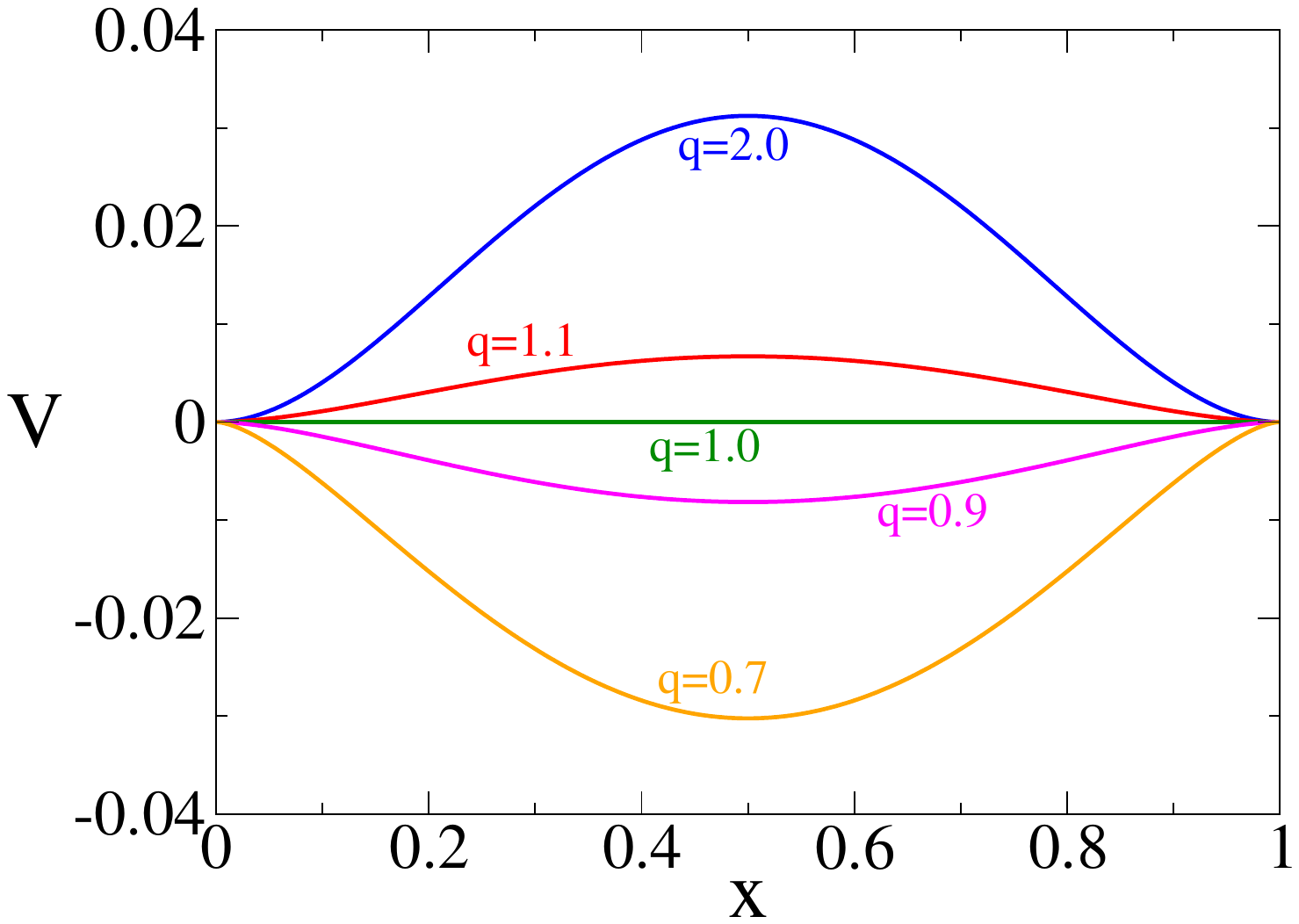}}
\caption{Upper panel: Illustration of the fixed-point structure of the rate equation (\ref{eq:rate_binary}) for the two-state NLVM, for $q<1$ and $q>1$ respectively. Full symbols represent stable fixed points, open symbols stand for unstable fixed points. Lower panel: Potential $V(x)$ in Eq.~(\ref{eq:V-x}), for different values of $q$. 
\label{fig:fp_m2_cg}  }
\end{figure}

We can now distinguish different regimes, depending on the value of the exponent $q$, as illustrated in Fig.~\ref{fig:fp_m2_cg}.

\begin{figure}
\centering 
\includegraphics[width=0.9\columnwidth]{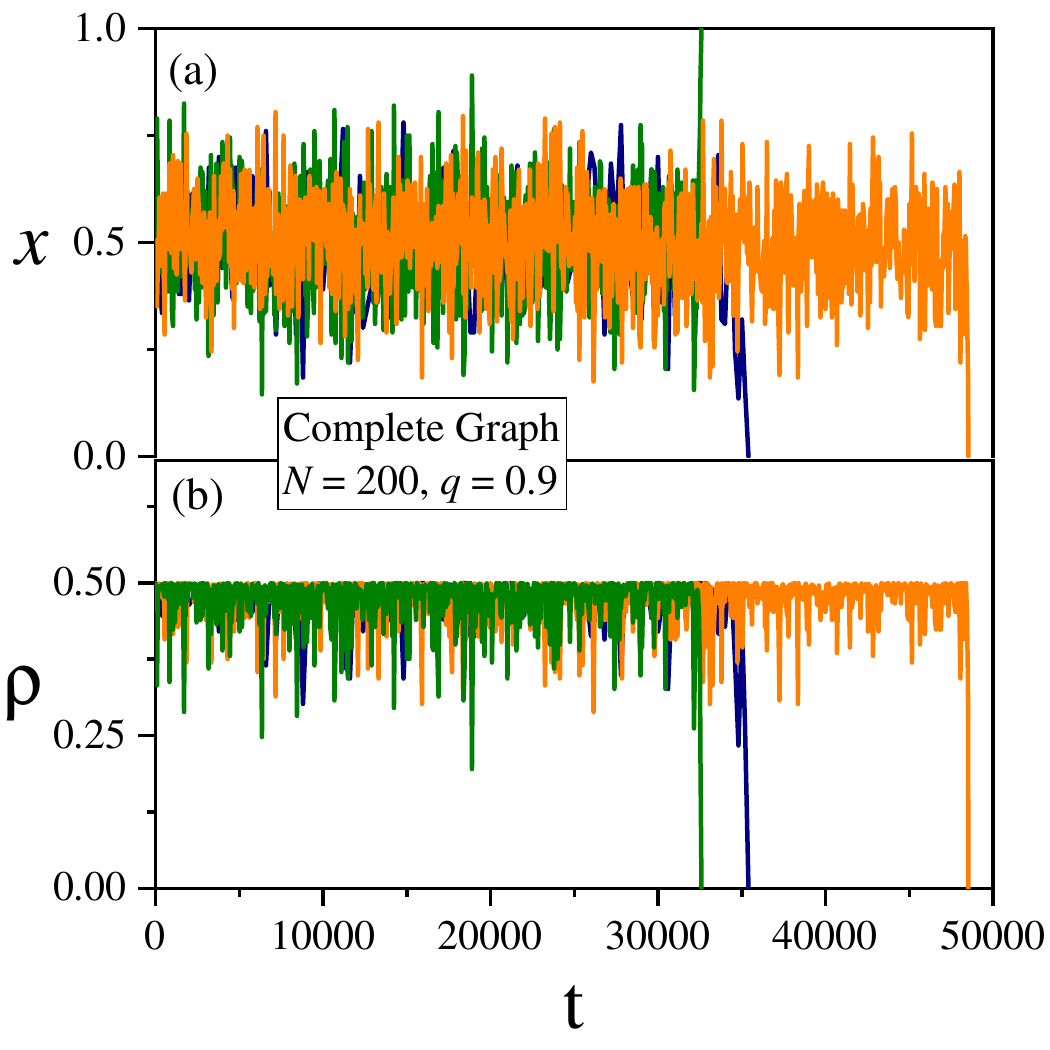}
\caption{Time evolution of individual realizations of the two-state ($M=2$) NLVM on complete graphs ($N=200$). Panel (a) shows different realizations for $q=0.9$. All runs are seen to remain near $x\approx 1/2$ for significant amounts of time until finite-size fluctuations take the system to consensus. Panel (b) shows the density of active links for the realisations in panel (a).
\label{fig:q_lt_1}
}
\end{figure}

For $q<1$ the central fixed point $x^*=1/2$ is linearly stable and a minimum of the potential. The remaining two fixed points are unstable (maxima of the potential). Infinite populations started from any $0<x<1$ will therefore converge to $x^*=1/2$, and remain there indefinitely. Finite populations first fluctuate around the deterministic fixed point $x^*=1/2$, and are then taken to consensus by a sudden finite-size fluctuation as illustrated in Fig.~\ref{fig:q_lt_1}. As we will discuss in more detail below the time to consensus increases exponentially with $N$ (Arrhenius law). We determine an approximate scaling of the mean consensus time in Sec.~\ref{sublinear}.

\begin{figure}
\centering
\includegraphics[width=0.9\columnwidth]{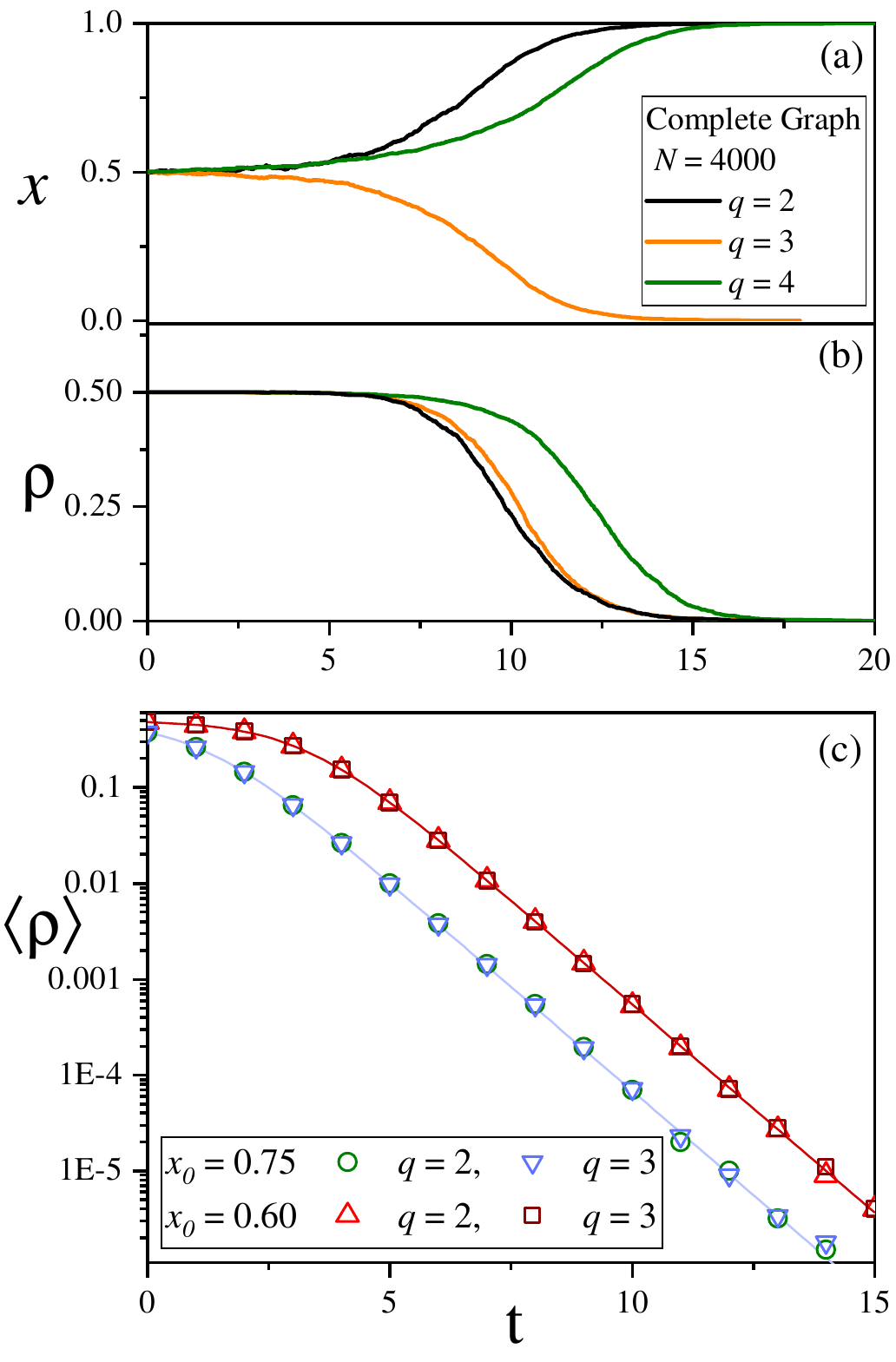}
\caption{Behavior of the two-state NLVM with $q>1$ on complete graphs. Panels (a), (b): Time evolution of the fraction of agents with opinion $1$, $x$, and the density of active links $\rho$, respectively, in individual realisations ($N=4000$).  Panel (c): Average density of active links, $\avg{\rho}$, in a system of size $N=10^4$ for $q=2$ and $3$, and different initial conditions, as indicated. Averages are over $10^4$ independent realizations. The solid curves are the analytical prediction from Eq.~(\ref{eq:rho-t-q2}). }
\label{fig:q_gt_1}
\end{figure}

For $q>1$ stability is reversed [Fig.~\ref{fig:fp_m2_cg}]. The central fixed point $x^*=1/2$ is unstable and a maximum of the potential, and the consensus states $x^*=0$ and $x^*=1$ are linearly stable minima. Deterministic drift thus pulls the system towards one of the absorbing states. Intrinsic fluctuations in finite populations are largely irrelevant in this case (except perhaps during the initial departure from region near the unstable fixed point), trajectories of the stochastic system approach consensus relatively quickly as shown in Fig.~\ref{fig:q_gt_1}. The mean time to reach consensus grows logarithmically with the population size $N$, as we will describe in more detail in Sec.~\ref{superlinear}.

We highlight the special case $q=1$, the model then reduces to the conventional (linear) two-state VM. The right-hand side (RHS) of Eq.~(\ref{eq:rate_binary}) vanishes in this case, and we have $\dot x=0$ and $V(x)\equiv 0$. Any initial state $x$ is marginally stable, leading to a conservation of the average magnetisation \cite{suchecki2005b}.

\subsection{Further analysis for $q=2$ and $q=3$}
\label{superlinear}

\subsubsection{Average density of active links}

We now study the special cases $q=2$ and $q=3$ respectively. Analytical progress is then possible.

For $q=2$, Eq.~(\ref{eq:rate_binary}) becomes
\be 
\dot {x}=x(1- x)(2x-1),
\label{eq:rate_binary-q2}
\ee
and the evolution of the density of active links $\rho=2x(1-x)$ fulfills
\be
\dot{\rho}=-\rho(1-2\rho).
\ee
The solution with initial condition $\rho(t=0)=\rho_0$ is
\be
\rho(t)=\frac{\rho_0}{2\rho_0+(1-2\rho_0)e^t}.
\label{eq:rho-t-q2}
\ee
Surprisingly, the rate equation for $x$ for $q=3$ turns out to be identical to that for the case $q=2$  [Eq.~(\ref{eq:rate_binary-q2})] and, therefore, the evolution of $\rho$ is also given by Eq.~(\ref{eq:rho-t-q2}) when $q=3$. 

Eq.~(\ref{eq:rho-t-q2}) indicates that the density of active links $\rho$ decays exponentially, except for the initial condition $x=1/2$ (leading to $\rho_0=1/2$). In this latter case $x$ and $\rho$ remain at their initial values indefinitely. This is a consequence of the symmetry between the two opinion states, and a valid description only on average (over realizations). In any single realisation of a finite system fluctuations will break the symmetry, and the system will eventually reach consensus. 

In Fig.~\ref{fig:q_gt_1}(c) we compare the  analytical approximation from Eq.~(\ref{eq:rho-t-q2}), derived in the limit $N\to\infty$, with the average value of $\rho$ over several realisations of the dynamics, for $q=2$ and $q=3$.


\subsubsection{Mean time to consensus}
Eq.~(\ref{eq:rho-t-q2}) can also be used to construct a rough estimate of the mean consensus time $\tau$. To do this, we assume that the system has reached consensus when either $x<1/N$ or $x>1-1/N$, that is the minority opinion is (on average) only represented by less than one individual. Using $\rho=2x(1-x)$ this translates into $\rho<\rho_1\equiv\frac{2}{N}(1-\frac{1}{N})$. The subscript $1$ indicates that this is the density of active links when all but one agent are in the same opinion state. 

We use the time $\tau$ at which the solution of Eq.~(\ref{eq:rho-t-q2}) takes the value $\rho_1$ as an estimate of the consensus time. We obtain
\be\label{eq:tau_estimate}
    \tau_1 = \ln \left[ \frac{\rho_0 N (N-4+4/N)}{2(1-2\rho_0)(N-1)} \right].
\ee
For $N\gg 1$ this can further be approximated, 
\be
    \tau_1 \approx \ln \left[ \frac{\rho_0 N}{2(1-2\rho_0)} \right].
    \label{eq:tau-N-q2}
\ee
This leads to $\tau_1\approx \ln\,N$ for large $N$.

\begin{figure}
\centering
{\includegraphics[width=0.85\columnwidth]{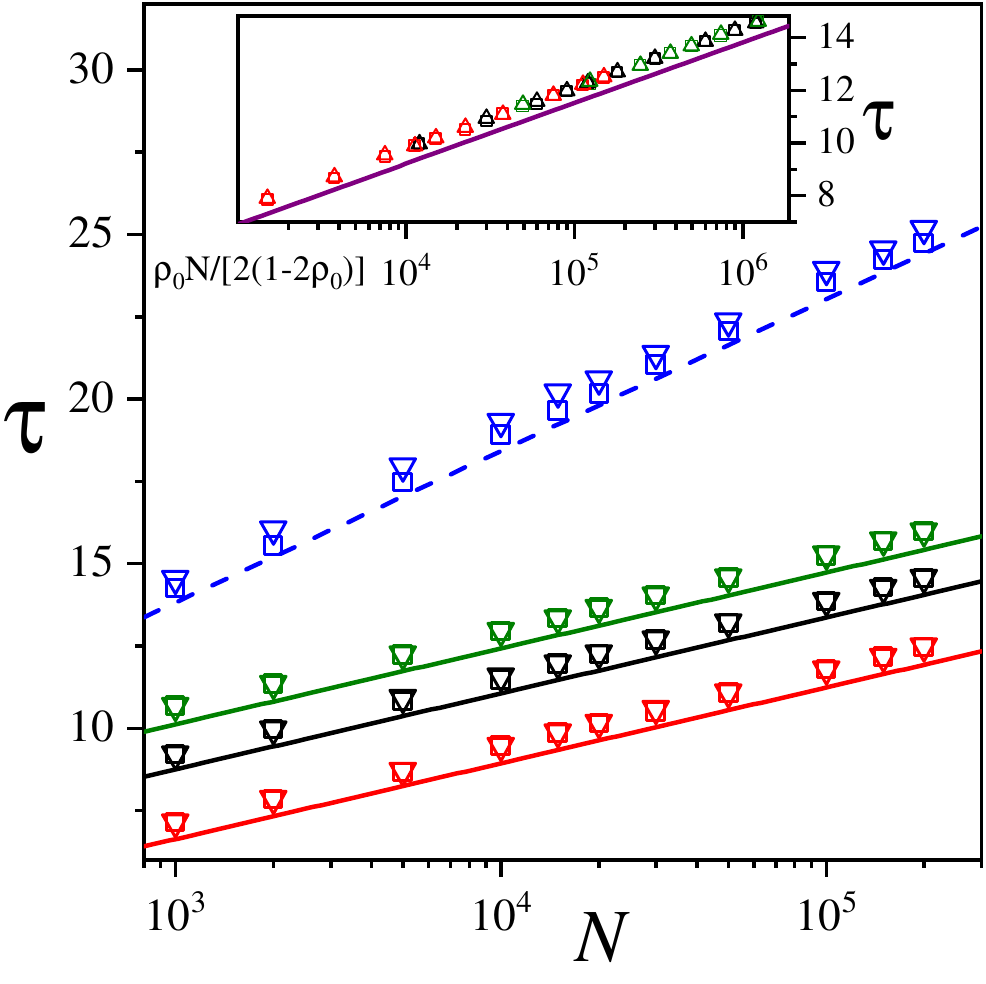}}
\caption{Mean consensus time $\tau$ vs system size $N$ for the $M=2$ NLVM on a CG, for $q=2$ (squares) and $q=3$ (triangles), and different initial conditions $x_0=0.5, 0.55, 0.60$ and $0.75$ (from top to bottom). Averages were obtained from $10^4$ independent realizations. Solid straight lines are the analytical approximation from Eq.~(\ref{eq:tau-N-q2}), while the dashed straight line shows the approximation $\tau \approx 2 \ln N$, valid for $x_0=1/2$. The slope for $x_0=0.5$ is twice as large as for the cases with $x_0 \ne 0.5$. Inset: Collapse plot for $x_0 \neq 0.5$. The horizontal axis has been scaled by the factor $\rho_0/2(1-2\rho_0)$.}  
\label{fig:tau-N-CG}
\end{figure}

In Fig.~\ref{fig:tau-N-CG} we plot $\tau$, as measured in simulations, as a function of $N$ for different initial conditions $x_0$. We find that the approximate expression from Eq.~(\ref{eq:tau-N-q2}) (solid lines) captures the behavior of $\tau$ with $N$ observed in simulations (symbols), and for various initial conditions $x_0 \ne 1/2$ (leading to $\rho_0<1/2$). We find that results from simulations are consistently higher by approximately $0.5$ units of time compared to the analytical estimate for all values of $N$ shown. We attribute this to the fact that some additional time is needed to reach consensus from the point at which the minority consists of one agent (we used the latter to obtain our analytical estimate of $\tau$). 

The initial condition $\rho_0=1/2$ ($x_0=1/2$) needs to be treated separately. In simulations we find the approximate scaling $\tau \sim 2 \ln N$ (blue symbols in Fig.~\ref{fig:tau-N-CG}), i.e., for large $N$ the consensus time with initial symmetric conditions is twice as large as that for an asymmetric starting point. A similar result was found for the majority rule model in Ref.~\cite{Redner-2003}.  This scaling can be found analytically (blue dashed line) by setting $\rho_0=1/2-2/N$ ($x=1/2 \pm 1/\sqrt{N}$) as the initial condition in Eq.~(\ref{eq:tau-N-q2}), leading to $\tau \sim 2 \ln\,N$ for $N \gg 1$.

We emphasize that for the present cases ($q=2$ and $q=3$) the initial escape is from the vicinity of a linearly unstable state $\rho_0=1/2$. Our results agree with a more general result for a decay from an unstable state in which $\tau \propto \ln \epsilon^{-1}$, where $\epsilon$ is the variance of the noise acting on the system \cite{QDT,toral_SM}. For the present model we have $\epsilon \propto 1/N$. 

\subsection{Quasi-linear dynamics ($q \gtrsim 1$)}

It is further interesting to analyze the model in the parameter region $q>1$, but very close to the linear case $q=1$. To do this we expand the RHS of Eq.~(\ref{eq:rate_binary}) in powers of $q-1$. To leading order the RHS is zero (conventional linear two-state VM), and to sub-leading order we find 
\be\label{eq:xdot_small_q-1}
    \dot{x} = (q-1) x(1-x) \ln \left( \frac{x}{1-x} \right).
\ee
From this one obtains
\be\label{eq:aux_sol}
    x(t) = \left[1+\left(\frac{1-x_0}{x_0}\right)^{e^{(q-1)t}}  \right]^{-1}
\ee
to be supplemented by an initial condition $x_0$ at time zero. For $q=1$ this reduces to $x(t)=x_0$ for all $t \ge 0$. 

For $q>1$ and $x_0>1/2$ the solution in Eq.~(\ref{eq:aux_sol}) approaches the asymptotic value $x=1$ as $t \to \infty$, albeit very slowly because $q-1$ is small. For $x_0<1/2$ the solution approaches $x=0$. 

Focusing on $x_0<1/2$ and assuming that the system reaches consensus at a time $\tau_1$ for which $x(\tau_1)=1/N$ we find the following estimate for the mean consensus time 
\be \label{eq:tau-q-sim-1}
    \tau_1 \approx \frac{1}{(q-1)} \ln \left[ \frac{\ln N}{\ln \left( \frac{1-x_0}{x_0}\right)} \right] ~~~~ \mbox{for $N \gg 1$}.
\ee
We recall that we have assumed that $q-1$ is positive but small to derive Eq.~(\ref{eq:xdot_small_q-1}).  

\begin{figure}[t]
\centering
\subfloat {\includegraphics[width=0.85\columnwidth]{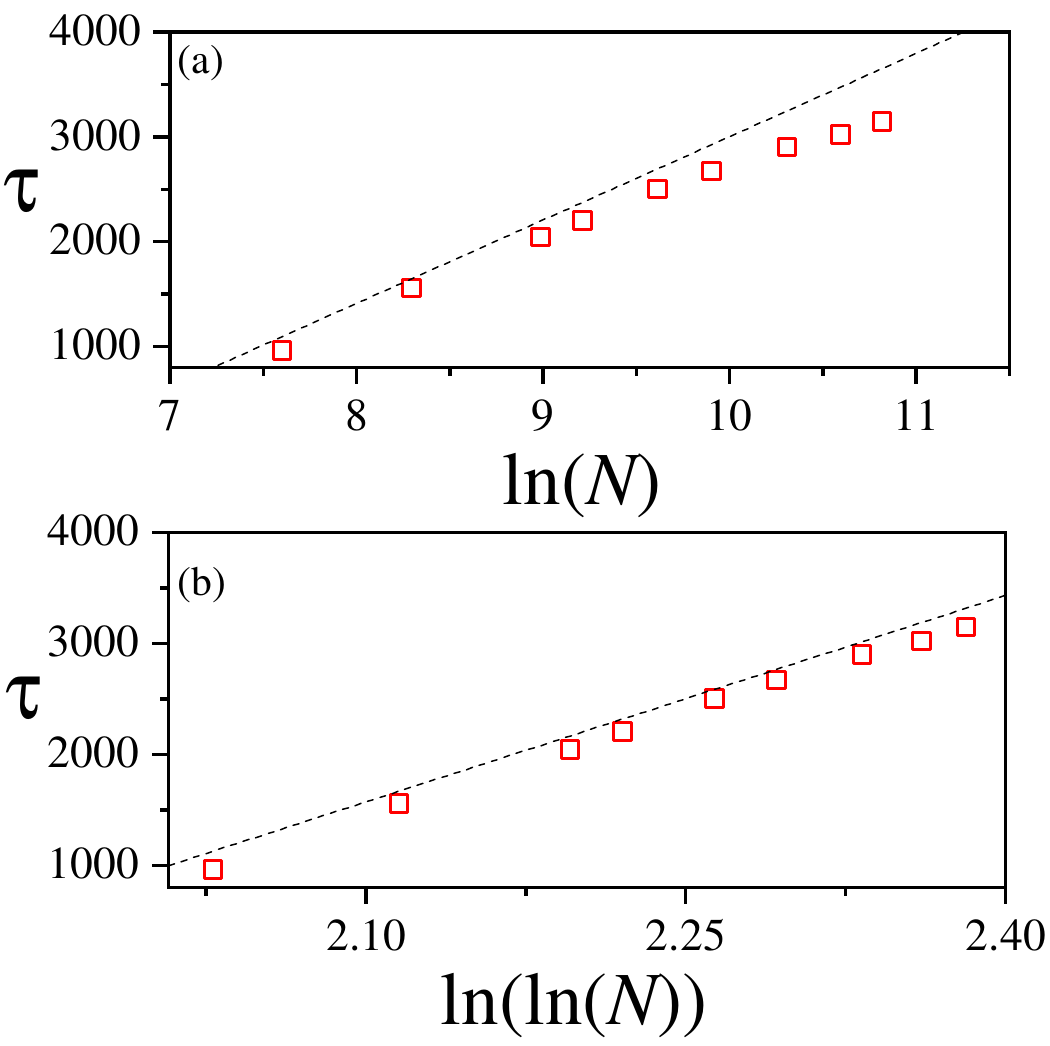}}
\caption{ Mean consensus time $\tau$ vs $\ln N$ [panel (a)] and vs $\ln(\ln N)$ [panel (b)]. Data is from simulations for $q=1.01$, starting from the initial condition $x_0=0.4$. Dotted lines are guides to the eye.}
\label{fig:tau-q-sim-1}
\end{figure}

We have investigated the scaling of the consensus time with $N$ in simulations, results are shown in Fig.~\ref{fig:tau-q-sim-1}. We do not suggest that the figure provides clear verification of the $\ln(\ln\,N)$ scaling predicted by Eq.~(\ref{eq:tau-q-sim-1}). Nonetheless we observe a slight downward bend in panel (a) where the horizontal axis shows $\ln(N)$. This bend is less pronounced in  panel (b), where $\tau$ is plotted as a function of $\ln(\ln\,N)$. This is an indication that the predicted $\ln(\ln\,N)$-scaling might be more consistent with the data than a simple $\ln\,N$ dependence.

\subsection{Sub-linear dynamics ($q < 1$)}
\label{sublinear}

When $q<1$ the two-state model quickly reaches a long-lived state in which densities fluctuate around $x=1/2$ (see Sec.~\ref{sec:rate_eq}). Consensus at $x=0$ or $x=1$ is eventually reached by finite-size fluctuations.  

An approximate expression for the behavior of the mean consensus time $\tau$ with $q$ and $N$ can be obtained from the Arrhenius formula, which predicts that the mean time for a stochastic process with noise of variance $D$ to overcome a potential barrier of height $\Delta V$ will be proportional to $e^{\Delta V/ D}$ (see for instance \cite{gardiner}).

Using this, the mean consensus time in the two-state NLVM for a fixed $q$ can be expected to be of the form $\tau=K e^{[V(x=0)-V(x=1/2)]/D}$,  where $K$ is a constant to be determined, and $V(x=0)-V(x=1/2)$ is the height of the barrier given by the potential $V(x)$ of Eq.~(\ref{eq:V-x}) [see Fig.~\ref{fig:fp_m2_cg}(c)].  The diffusion coefficient $D$ can be obtained from the Fokker-Planck equation for the model, which in turn can be derived from the master equation~(\ref{eq:min_generalM}) for $M=2$ by means of a Kramers--Moyal expansion in the limit of large but finite $N$ \cite{gardiner}. We find
 \be \label{eq:FP}
\frac{\partial P(x, t)}{\partial t}= -  \frac{\partial(\mathcal{F}(x)P(x, t))}{\partial x} + \frac{\partial(\mathcal{D}(x)P(x, t))}{\partial  x^2},
\ee
where $P(x, t)$ is the probability distribution for the model, written in terms of $x$. The functions $\mathcal{F}(x)$ and $\mathcal{D}(x)$ are the  drift and diffusion terms, respectively. Those terms are given by the transition rates in Eq.~(\ref{eq:rate_cg_m2})  as $\mathcal{F}(x) = [{\cal T}^{-} - {\cal T}^{+}] /N $ and $\mathcal{D}(x) = [{\cal T}^{-}+ {\cal T}^{+}] /(2N^2)$, where the rates ${\cal T}^\pm$ are to be evaluated at the point $n=xN$. We find
\be \label{eq:F}
\mathcal{F}(x) =  x(1-x)^q-(1-x)x^q,
\ee
and
 \be \label{eq:D}
     {\cal D}(x)= \frac{(1-x)x^q+x(1-x)^q}{2N}.
\ee
The drift term ${\cal F}(x)$ is also the RHS of Eq.~(\ref{eq:rate_binary}).

\begin{figure}[t]
\centering
\includegraphics[width=0.85\columnwidth]{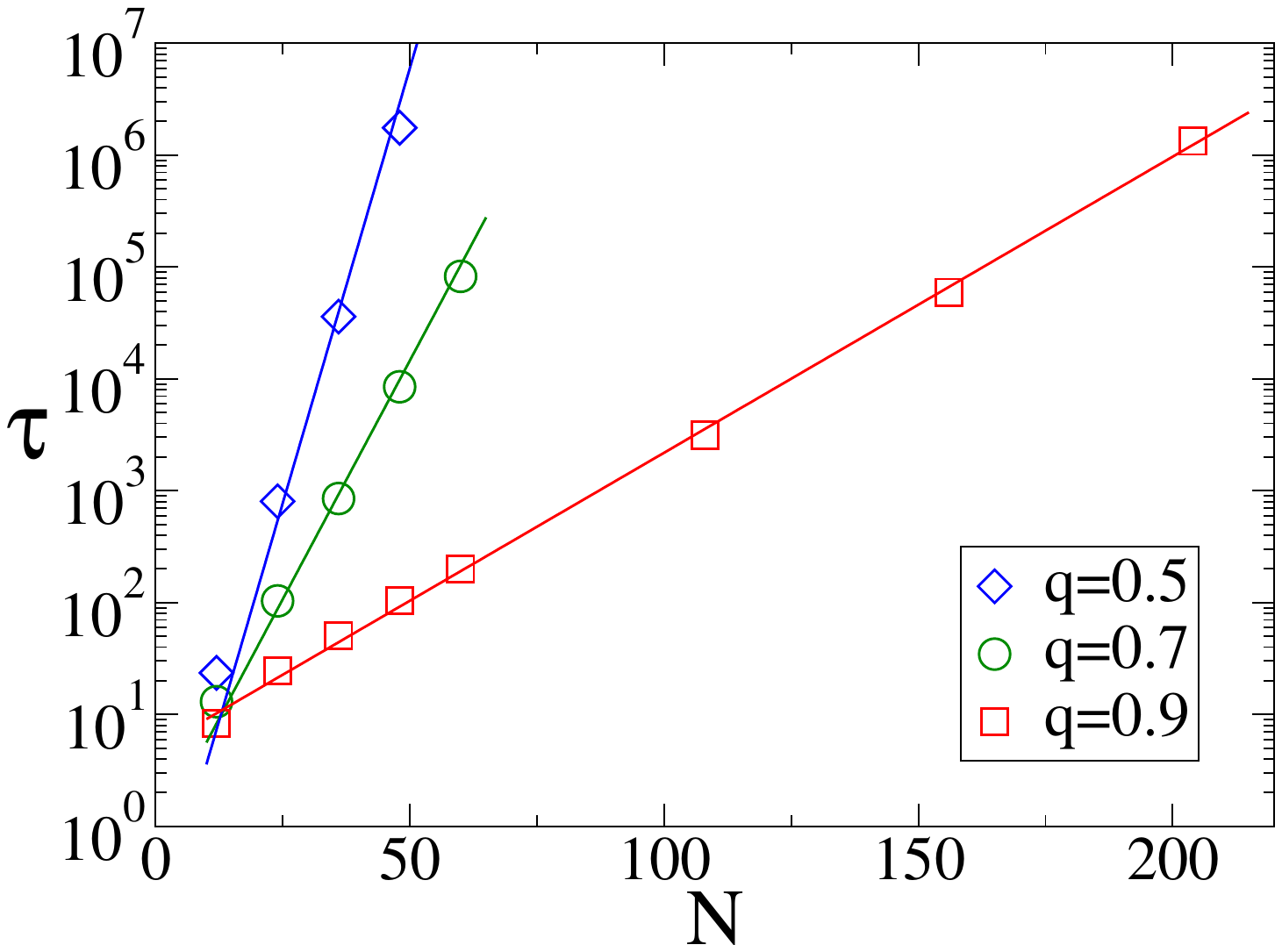}
\caption{Mean consensus time $\tau$ vs system size $N$ for the two-state NLVM on a complete graph. Averages are over $10^3$ realizations, starting from the initial condition $x_0=0.5$. The solid lines are from the analytical approximation Eq.~(\ref{eq:tau-N-q}) adjusted by a factor $K(q)$ to fit the data. }
\label{fig:consenus-q}
\end{figure}

We note that the Arrhenius formula is valid for a diffusion coefficient $D$ that is independent of $x$, while for the present model the diffusion coefficient resulting from a Kramers--Moyal expansion depends on the state of the system. We use the constant value $D= \frac{1}{2}[\mathcal D(0)+\mathcal D(1/2)]=\left[ 2^{(q+2)}N \right]^{-1}$ as an approximation in the Arrhenius formula. Further, from Eq.~(\ref{eq:V-x}) we obtain
\be
    \Delta V = V(0)-V(1/2) = \frac{\left[ 1-2^q \right]}{(q+1)2^q} - \frac{\left[ 1-2^{(q+1)} \right]}{(q+2)2^{(q+1)}}.
\ee
 Putting everything together, we find the approximation
\be
    \tau= K(q)\exp{ \Biggl\{ 2 \biggl\{ \frac{2\left[ 1-2^q \right]}{(q+1)} - \frac{\left[ 1-2^{(q+1)} \right]}{(q+2)}} \biggr\} N \Biggr\},
    \label{eq:tau-N-q}
\ee
where $K(q)$ is a pre-factor, which would need to be determined separately. This result is tested against simulations in Fig.~\ref{fig:consenus-q}, where we see that Eq.~(\ref{eq:tau-N-q}) captures the right scaling behavior of $\tau$ with $N$ for different values of $q<1$. 

If we assume that the transition from the metastable state at $q<1$ to consensus is a Poisson process taking place with rate $1/\tau$, and taking into account that $\rho\approx 1/2$ in the metastable state, then we expect that the ensemble-averaged density of active links follows the exponential decay
\begin{equation}
    \langle \rho \rangle(t) \approx \frac{1}{2} e^{-t/\tau}.        
    \label{eq:rho-ave-t}
\end{equation}

\begin{figure}[t]
\centering
\subfloat {\includegraphics[width=0.85\columnwidth]{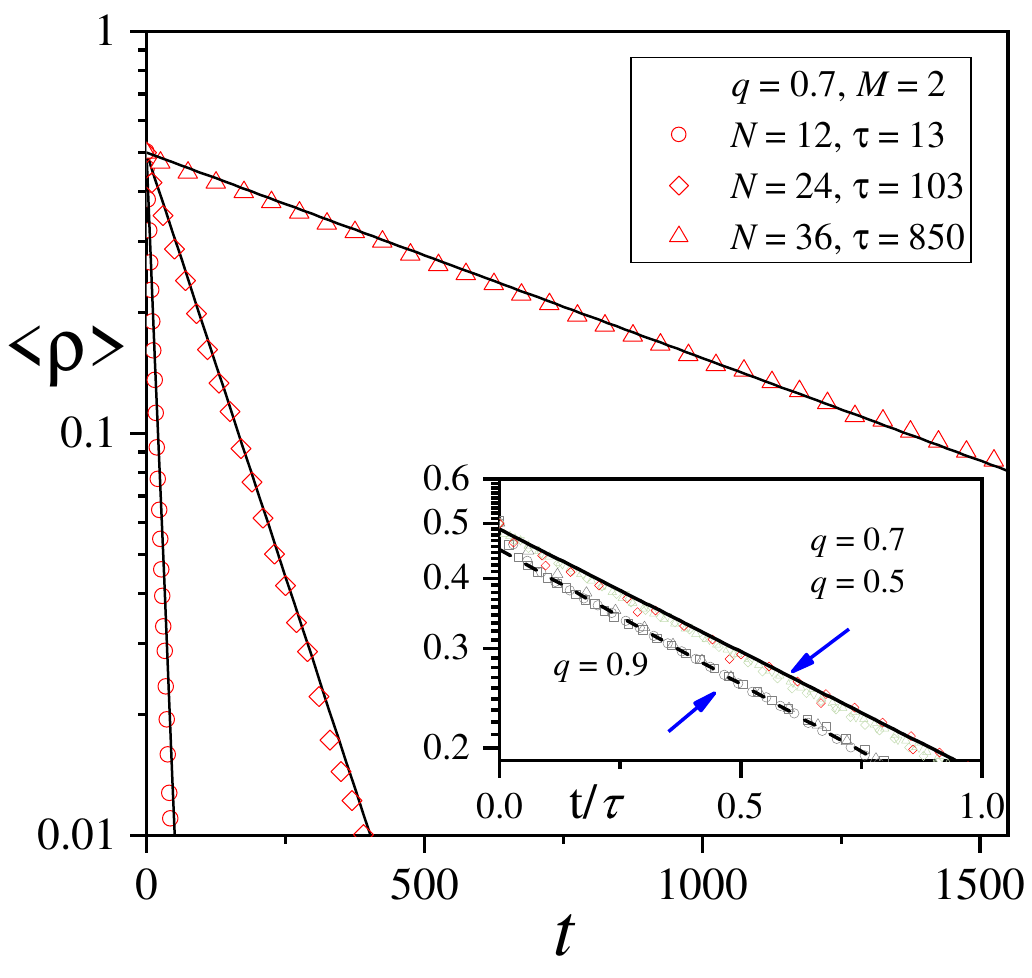}}
\caption{Time evolution of the ensemble averaged density of active links in the two-state model on complete graphs, for $q = 0.7$.  Symbols are from simulations, lines are the approximation in Eq.~(\ref{eq:rho-ave-t}), with the numerical values of $\tau$ obtained from simulations (see Fig.~\ref{fig:consenus-q}). The inset shows the same data, but plotted as a function of $t/\tau$. The full line is the approximation from Eq.~(\ref{eq:rho-ave-t}), while the dashed line is an exponential fit for the case $q=0.9$, and proportional to $e^{-t/(0.9\tau)}$.}
\label{fig:rho-q-less-1}
\end{figure}

In the main panel of Fig.~\ref{fig:rho-q-less-1} we plot $\langle \rho (t) \rangle$ for $q=0.7$ and three different system sizes.  We see the approximation from Eq.~(\ref{eq:rho-ave-t}) fits the data quite well.  The inset shows $\rho$ for $q=0.5$, $0.7$ and $0.9$ plotted against $t/\tau$. We find good collapse of the curves for $q=0.5$ and $0.7$, while there is a small deviation in the slope for $q=0.9$. This discrepancy is likely due to the fact that the potential $V(x)$ becomes increasingly more shallow near $q=1$, so that the Arrhenius law no longer applies.

In summary, while the true stationary solution of the two-state models consists of two delta peaks at the absorbing states, we find long-lived metastable states for $q<1$, with a lifetime that grows exponentially with $N$. These long-lived states are centred on $x=0.5$, and fluctuations about this point can be described by a quasistationary distribution (QSD) \cite{ing,qst,jaume}. This distribution characterizes the system on the time scale given by Eq.~(\ref{eq:tau-N-q}). 
 
\begin{figure}
\centering
\includegraphics[width=0.85\columnwidth]{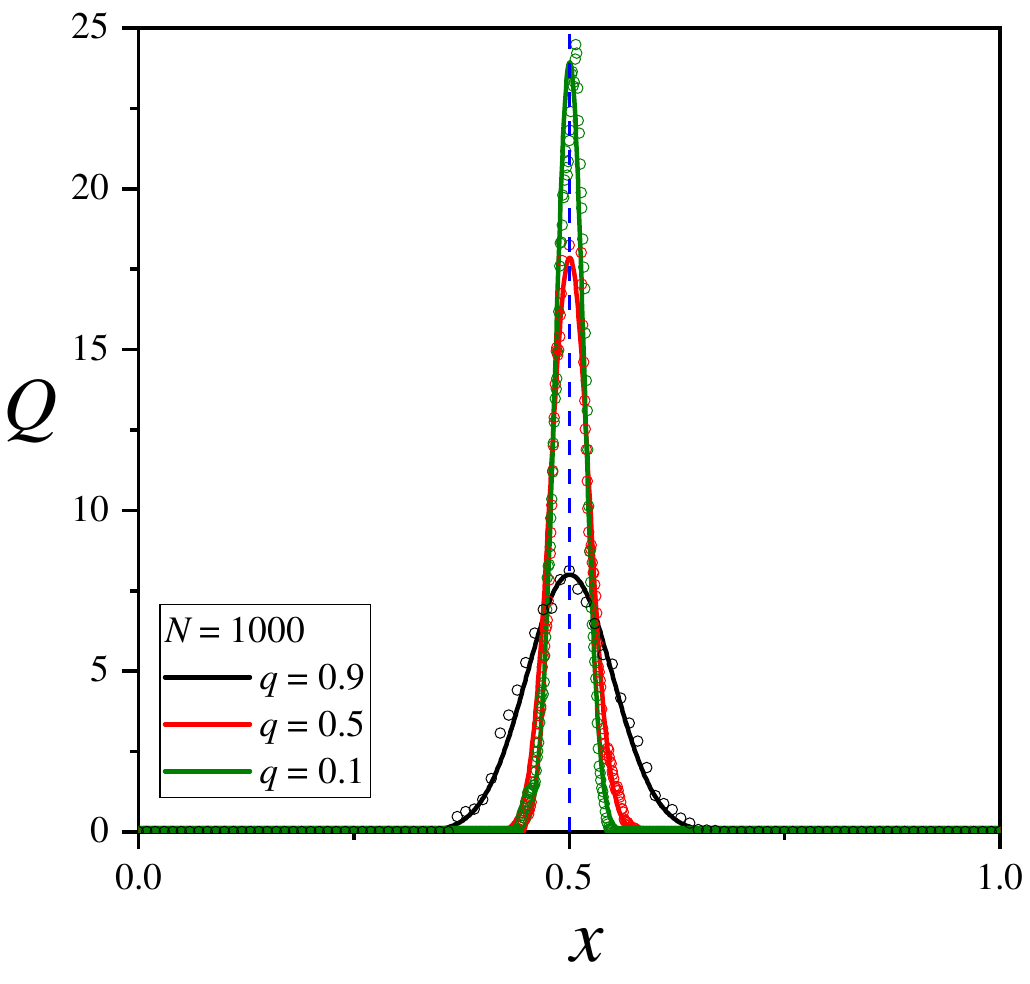}
\caption{Quasi-stationary probability distribution for the two-state NLVM on a complete graph ($N=100$), for different values $q<1$. Full lines correspond to the QSD obtained from the numerical integration of a modified master equation ~\cite{ing,jaume}. Symbols are from simulations.}
\label{fig:quasi}
\end{figure}

 In terms of the original variable $n$, one can define the distribution $Q(n,t)=P(n,t |0<n<N)$, that is the distribution of $n$ at time $t$, conditioned on the fact that the process has not reached absorption by that time. The QSD, $Q(n)$ is the long-time limit of $Q(n,t)$, and can be obtained as the stationary solution of a modified master equation~\cite{ing,jaume}.  In simulations, the QSD can be measured by averaging only over realisations that have not reached the absorbing states. 
Fig.~\ref{fig:quasi} shows results for the QSD for the two-state model on complete graphs and for different values of the parameter $q$. The calculation of the QSD from a master equation \cite{jaume} gives an overall good description of the results form numerical simulations.

\begin{figure*}[t]
\centering
\includegraphics[width=0.8\textwidth]{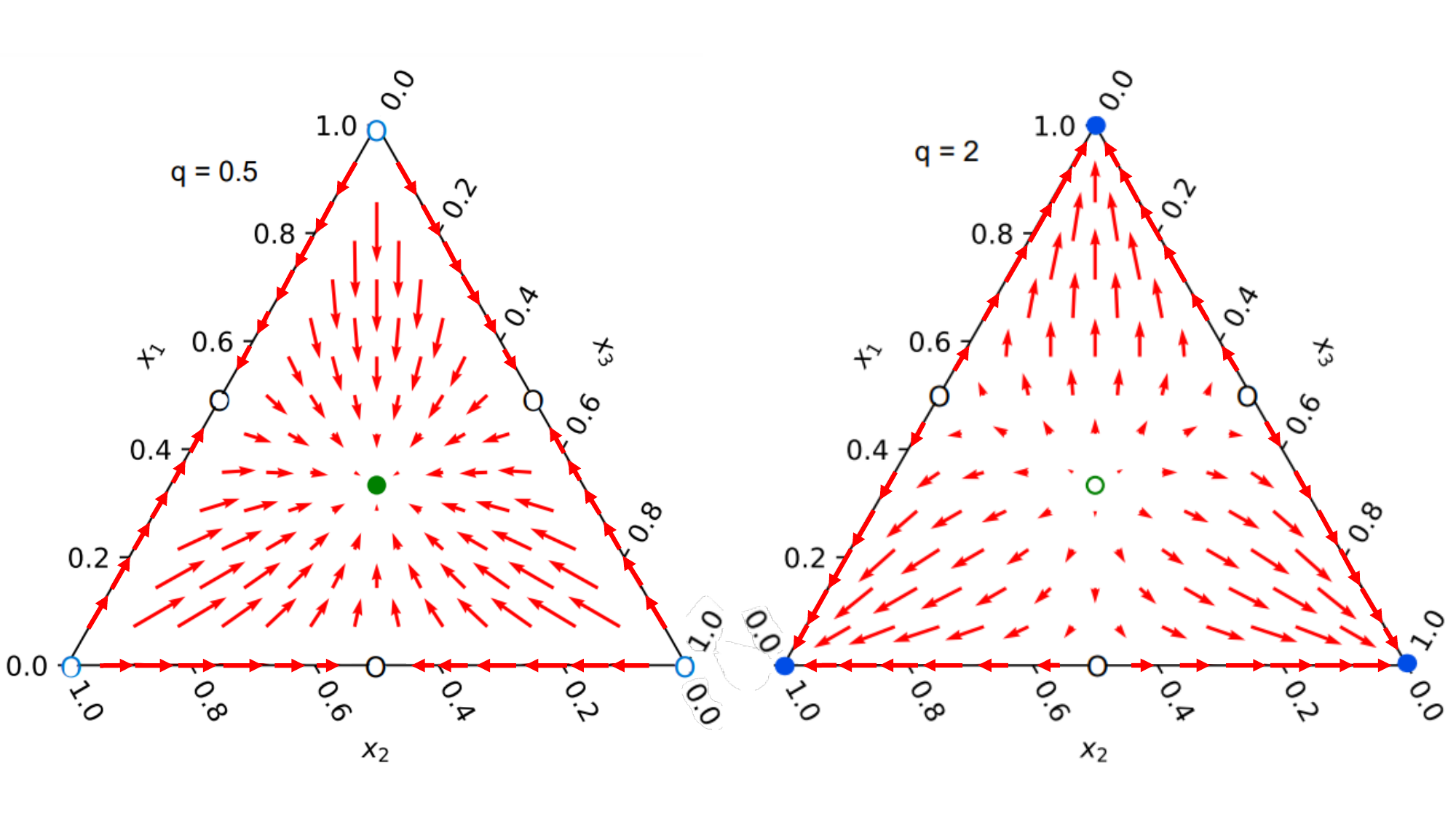}
\caption{Flow diagram of the system dynamics for the nonlinear VM with $M=3$ states in a ternary plot. Arrows represent the direction of the deterministic flow [Eqs.~(\ref{eq:rateM})]. Fixed points are represented by full circles (stable in the phase space of the full model), and empty circles (unstable) respectively. Black empty circles indicate saddle points.  \label{fig:fpcgm3}}
\end{figure*}
\section{Multistate NLVM model on complete graphs}\label{sec:multi_state}

\subsection{Reduction to an effective two-state model is not possible for $q\neq 1$}
The dynamics of the linear model ($q=1$) with $M$ opinion states can, for many purposes, be reduced to that of a two-state model, see e.g. \cite{redner, herrerias, msvm}. Similar ideas are also used in population genetics \cite{kimura,littler}. To carry out the reduction one focuses on one opinion state $\alpha$.  The rate at with which individuals switch out of this state on a complete graph is then given by 
\be
\label{eq:rate_lineal_msvm}
T_{\alpha,-}^{\rm eff}=\sum_{\beta\neq\alpha}n_\alpha\frac{n_{\beta}}{N-1}=n_\alpha\frac{N-n_\alpha}{N-1},
\ee
The rate of individuals changing into state $\alpha$ is given by the same expression. This results in an effective model distinguishing only between two types of agents, those in state $\alpha$ and those in any state that is different from $\alpha$. The resulting two-state model can then be used to predict the dynamics until the extinction of the focal opinion. 

On the other hand, the dynamical rules in Eq.~(\ref{eq:rate}) do not allow the same reduction for the NLVM. This can be seen from Eq.~(\ref{eq:rates_aa}). The total rate of leaving state $\alpha$ in the nonlinear model is given by
\be
T_{\alpha,-}=\sum_{\beta\neq\alpha}n_\alpha\left(\frac{ n_\beta}{N-1}\right)^q,
\ee
and it is clear that for $q\neq 1$ this cannot be written as a function of $n_\alpha$ only. Instead the full knowledge of all $\{n_\beta\}$ is required.

\subsection{Rate equations, fixed points and their stability}
 We write $x_\alpha=n_{\alpha}/N$ for the fraction of agents in opinion state $\alpha$, and $\mathbf{x}=(x_1,\dots,x_M)$. The rate equations then become

\be \label{eq:rateM}
\dot x_\alpha=(1- x_\alpha)x_\alpha^q-x_\alpha\left(\sum_{\beta \neq \alpha}^{M}{x_\beta}^q\right)
\ee
This is a set of $M$ coupled ordinary differential equations, and we note the reduction to $\dot x_\alpha=0$ for $q=1$. 

For general values of $q$, and a given value of $M$, the fixed points are of three different types.
First, there is the `full-coexistence point' of the model with $M$ opinion states,
\be\label{eq:xm}
\bx_M=\left(\frac{1}{M},\frac{1}{M},...,\frac{1}{M},\frac{1}{M}\right).
\ee
All opinions are present in equal proportions at this fixed point. 

Second, there are $M$ distinct `consensus fixed points' given by $(1,0,\dots,0), (0,1,0,\dots,0),\dots, (0,0,\dots, 1)$. At each of these only one single opinion remains.

In-between these, there are what we will refer to as the `partial-coexistence fixed points'. For each $\ell=2,\dots,M-1$ there are $M!/[(M-\ell)!\ell!]$ fixed points for which $\ell$ opinions are present with fractions $1/\ell$, and where the remaining $M-\ell$ opinions have gone extinct. The case $\ell=M-1$ for example describes fixed points of the form
\be\label{eq:xm-1}
\bx_{M-1}^{(\alpha)}=\left(\frac{1}{M-1},\frac{1}{M-1},...,0,\dots,\frac{1}{M-1}\right),
\ee
where the zero entry is in position $\alpha$ ($\alpha\in\{1,\dots,M\}$).

 Linear stability analysis shows that the full coexistence point $\bx_M$ is stable if and only if $q<1$.  The consensus states on the other hand are linearly stable if and only if $q>1$. 

The stability of the partial-coexistence fixed points is more intricate. These are fixed points at which some opinions have gone extinct, but where at least two opinions remain. In the full phase space of the model with $M$ opinions (the simplex defined by $x_1+\dots+x_M=1$) these fixed points are always saddle points for any $q<1$ or $q>1$. It is important to realise though that these fixed points are located at the (hyper) faces of the full simplex, and the faces are invariant under the dynamics in Eqs.~(\ref{eq:rateM}) (any $x_\alpha$ that is zero initially will remain zero). This type of solution is also found for the noisy nonlinear MSVM for integer $q$, when the noise rate is set to zero  \cite{sznajd_qmsvm}, and in majority-rule models with multiple states \cite{msvm-majority}. 

For $q<1$ the flow of Eqs.~(\ref{eq:rateM}) is directed away from the faces (or edges) of the full simplex. However, we note that each partial-coexistence fixed point in the model with $M$ opinion states is a full coexistence fixed point in a model with a lower number of opinion states. For example, the fixed points $x_{M-1}^{(\alpha)}$ in Eq.~(\ref{eq:xm-1}) represent full coexistence (at equal proportions) in a model with $M-1$ opinion states. For $q<1$ these fixed points are stable in the phase space of the reduced model.

For $q>1$ the full coexistence fixed point is linearly unstable, and within each face or edge the partial-coexistence fixed points are also unstable. The deterministic flow is directed towards the consensus states. 

We illustrate the different types of deterministic flow for the model with $M=3$ opinions in Fig.~\ref{fig:fpcgm3}. The full phase space is the simplex shown in the figure. The full coexistence fixed point is located in the centre. There are three partial-coexistence fixed points, one in the centre of each edge of the simplex. At these fixed points the population is divided equally between two opinions. Finally, there are the three consensus states at the corners of the simplex.

For $q<1$ the central full co-existence fixed point is linearly stable, see panel (a). The flow of Eqs.~(\ref{eq:rateM}) is away from the edges of the simplex, but if the dynamics is restricted to an edge, then the corresponding partial coexistence fixed point is stable. The direction of the deterministic flow is reversed for $q>1$, see panel (b).

\begin{figure*}
\centering
\includegraphics[width=2\columnwidth]{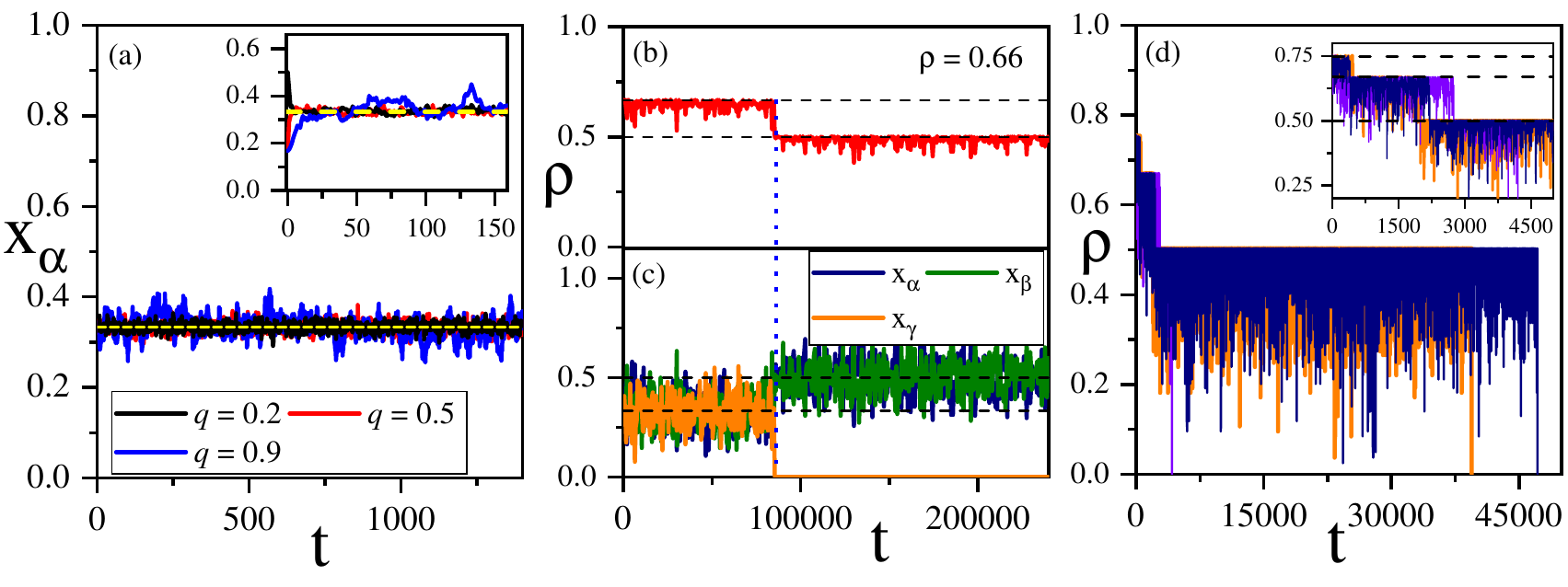}
\caption{Single realizations of the NLVM with $q<1$ on a CG. (a) Time evolution of the fraction of agents  $x_{\alpha}$ in a population of $N=3000$ agents and $M=3$ opinions (balanced initial conditions). The dashed yellow line shows $x_\alpha \equiv 1/3$.  Inset: Realisations with different initial conditions evolve to $x_\alpha \equiv 1/3$. (b) Evolution of $\rho$ for $q = 0.9$, $N=1000$ and $M=3$ (balanced initial conditions).  (c) Fractions $x_{\alpha}$ for the same realisation as in panel (b). Initially $\rho \simeq 0.66$ (full-coexistence fixed point of the $M=3$ model), until one opinion becomes extinct at time $t\approx 83000$ (vertical dashed line), and thus $\rho$ drops to a secondary plateau at $\rho \simeq 0.5$, corresponding to the full-coexistence fixed point of a model with two opinions. (d) Different realizations for the evolution of $\rho$ ($q=0.9$, $N=600$, $M=4$, balanced initial conditions). In each realisation, $\rho$ drops from $\approx 0.75$ to $0.66$ to $0.5$ to $0$, as opinions go extinct. } 
\label{fig:rho_qsmaller1_cg}
\end{figure*}



\subsection{Density of active links and dynamics in finite populations}

The density of active links [Eq.~(\ref{eq:defrho})] in the model on a complete graph is given by
\be
\rho = 2 \sum_{\alpha=1}^{M} \sum_{\beta>\alpha}^M x_\alpha x_\beta.
\ee
At the full-coexistence fixed point we thus have $\rho = (M-1)/M$. At the partial coexistence fixed points with $\ell$ remaining opinions, $\rho = (\ell-1)/\ell$. At the consensus fixed points there are no active links, $\rho=0$. 

We now briefly describe the behaviour of the density of active links in finite systems at the level of individual realizations (see Figs.~\ref{fig:1realisations_cg} and ~\ref{fig:rho_qsmaller1_cg}).

When $q<1$, the density of active links is first seen to fluctuate around the co-existence value $\rho = (M-1)/M$. Through finite-size fluctuation one opinion then becomes extinct. This then leads to a model with $M-1$ opinions, and the system settles near the full co-existence fixed point of this reduced model, i.e., around $\rho=(M-2)/(M-1)$. Then another extinction occurs, and so on. This leads to a succession of plateaux located at $\rho = (M-\ell-1)/(M-\ell)$, where $\ell$ represents the number of extinctions, until one opinion is left after $M-1$ extinctions.  This is illustrated in Figs.~\ref{fig:1realisations_cg}  and \ref{fig:rho_qsmaller1_cg} for $M = 3$ and $M = 4$. We note that the behavior is different from that of the linear multistate VM, where the series of plateaux can be quite different across realisations \cite{msvm}. This is a consequence of the deterministic pull towards the fixed points of the type $\rho=(M-\ell-1)/(M-\ell)$ in the nonlinear model with $q<1$ and $M-\ell$ remaining opinions. This pull is absent in the linear model.

For $q>1$ the behaviour of the multistate NLVM is very different to that for $q<1$. The deterministic flow is towards the consensus states and the density of active links in finite populations rapidly approaches $\rho=0$, as shown in Fig.~\ref{fig:1realisations_cg}(c). 
 
\subsection{Time evolution of the ensemble average of the density of active links}
We now turn to the behavior of ensemble-averaged density of active interfaces.
\subsubsection{$q<1$}
\begin{figure}
\includegraphics[width=0.85\columnwidth]{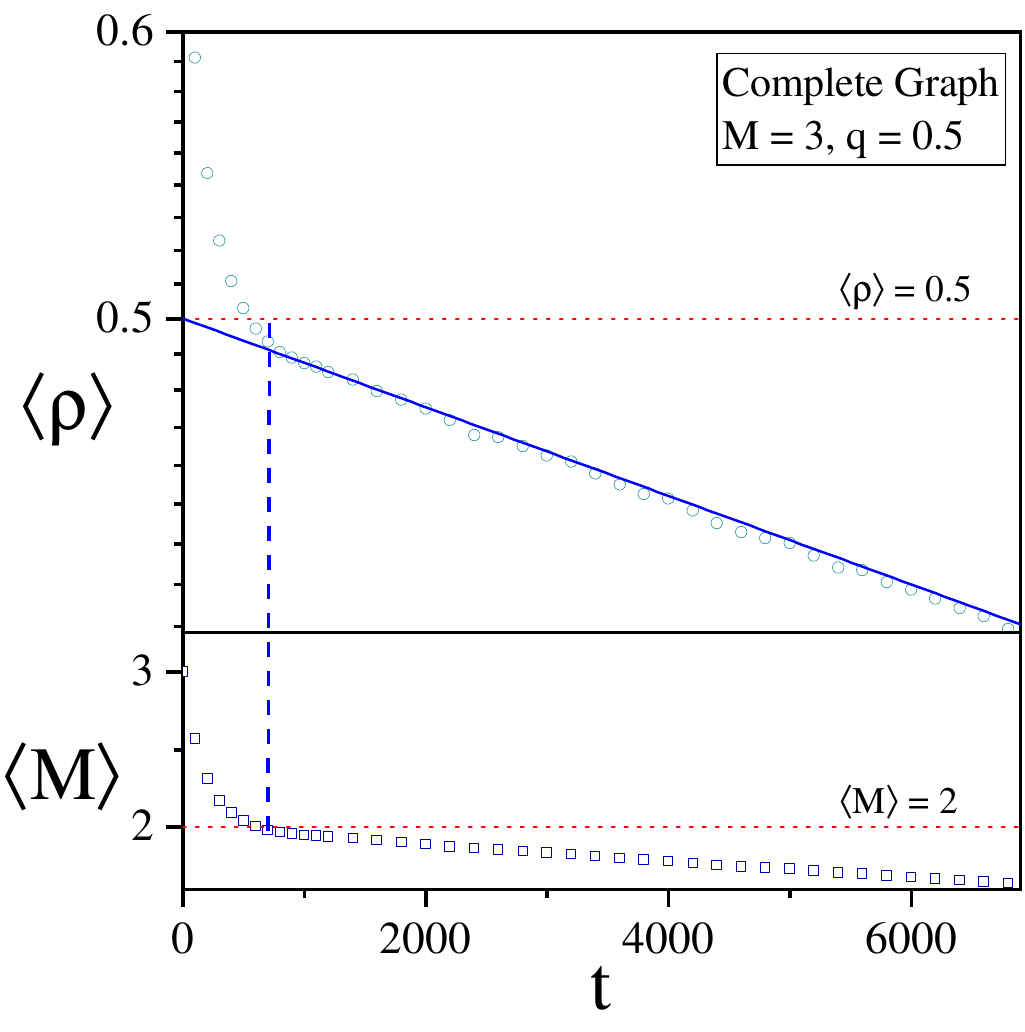}
\caption{Time evolution of the average density of active links (upper panel) and the mean number of opinion states present (lower panel) for the model with initially $M=3$ on a complete graph ($N = 36$, $q = 0.5$, average over $5000$ runs). The dynamics is started from balanced initial conditions. Symbols are from simulations. The blue line in the upper panel is from Eq.~(\ref{eq:rho-ave-t}) with $\tau$ from Eq.~(\ref{eq:tau-N-q}). As seen by comparing the upper and lower panels, the decay of $\avg{\rho}$ becomes exponential broadly from the time when, on average, there are only two opinions present.  
 \label{fig:decayMSVM}}
\end{figure}

We first focus on the model with $q<1$.  Fig.~\ref{fig:decayMSVM} shows the typical behavior of the average interface density, obtained from simulations. While the decay of $\avg{\rho(t)}$  is not exponential in the early phases, the interface density becomes exponential in the final stages when there are only at most two opinions left in the system. $\avg{\rho(t)}$ then follows the expression in Eq.~(\ref{eq:rho-ave-t}) for the nonlinear VM with only two opinions.

In the earlier stages of the time-evolution the system is characterised by a sequence of decay processes, first from $M$ opinions to $M-1$, then to $M-2$, and so on. We now study the dynamics in these different stages. In order to delineate the different stages, we adopt a basic heuristic. We measure, at each time, the average number of opinions still present in the system. We then record the times $t_\ell$ at which the average number of alive opinions is $\ell$ ($\ell=M-1, M-2,\dots,2$). Then we say that, broadly, that the decay from $M$ to $M-1$ opinions happens between time zero and time $t_{M-1}$, the decay from $M-1$ to $M-2$ opinions in the time interval from $t_{M-1}$ to $t_{M-2}$ and so on. We also record the average density of interfaces $\avg{\rho}_\ell\equiv \avg{\rho(t_\ell)}$ at each of the times $t_\ell$. 

\begin{figure}
\includegraphics[width=1\columnwidth]{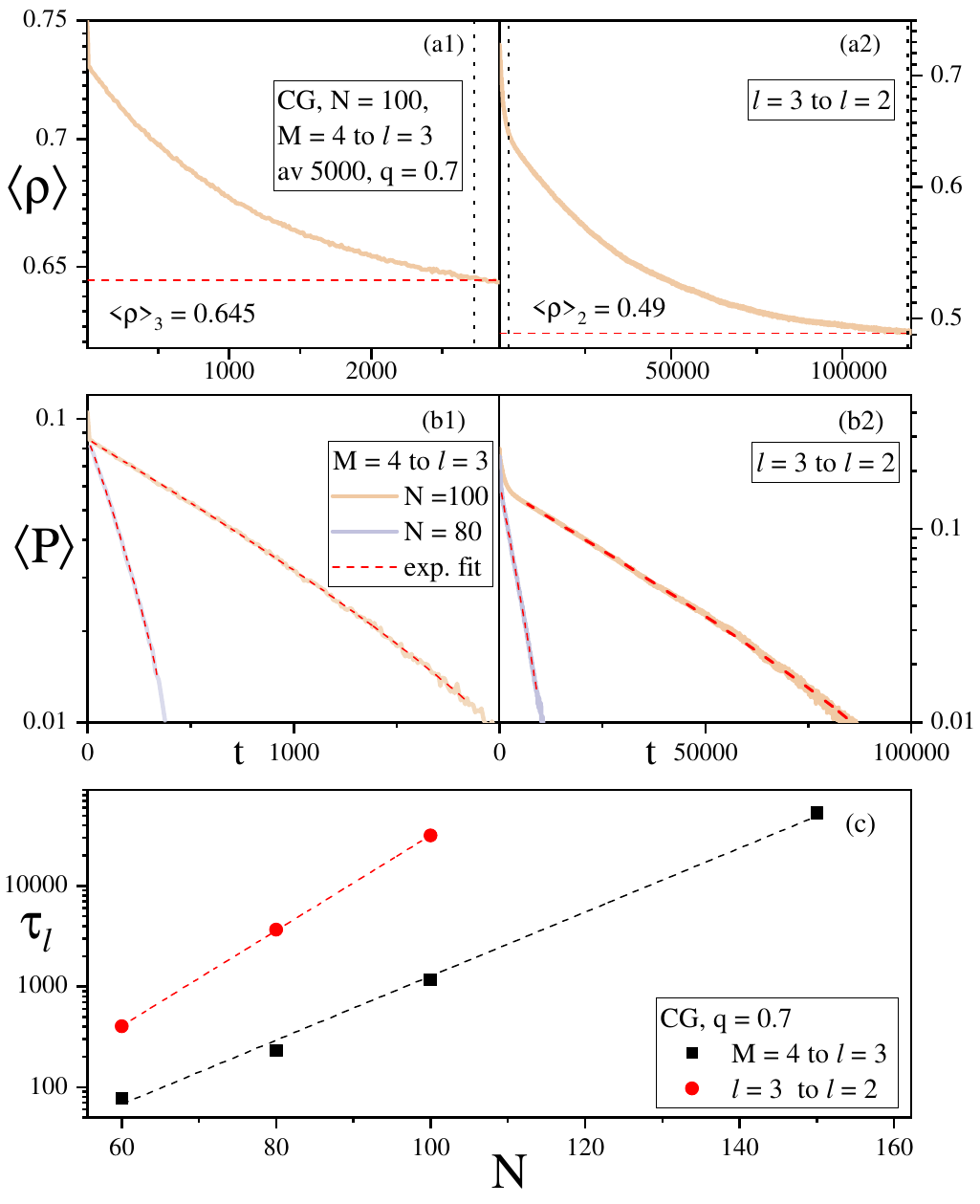}
\caption{(a) Time evolution of the average density of active links for the NLVM with $M=4$ opinions on a complete graph ($5000, N=100, q=0.7$).  Vertical dotted lines denote the time $t_3$ in panel (a1), and $t_2$ in (a2). Panel (b1) shows $\langle P \rangle \equiv \langle \rho \rangle - \langle \rho \rangle_3$ vs $t$, and (b2) $\langle P \rangle \equiv \langle \rho \rangle - \langle \rho \rangle_2$ vs $t$, for $N=100$ and $N=80$.  Dashed red lines are fits of the form $C e^{-t/\tau_l}$. The time constant $\tau_l$ is different in (b1) and (b2) respectively, and depends on $N$. (c) $\tau_l$ vs $N$ for the transition from $4$ to $3$ opinions (black squares), and for the transition from $3$ to $2$ opinions (red circles).  
\label{fig:scalingMSVM}}
\end{figure}

An example can be found in Fig.~\ref{fig:scalingMSVM}~(a1) and (a2), where we show the average density of interfaces for a model with $M=4$ during the time interval from $t=0$ to $t=t_2$ [panel (a1)], and in the time interval from $t_3$ to $t_2$ [panel (a1)]. The values of $\avg{\rho}_3$ and $\avg{\rho}_2$ are also indicated.

In Fig.~\ref{fig:scalingMSVM}(b) we then plot the quantity $\avg{P(t)}$, where $P(t)=\rho(t)- \avg{\rho}_\ell$. This quantity is seen to decay exponentially, confirming that the earlier stages of the time-evolution can also be seen as Poissonian decay processes. The corresponding time scales $\tau_3$ and $\tau_2$ can be extracted from exponential fits to the data in panels (b1) and (b2), respectively. We highlight that $\tau_\ell$ grows exponentially with the system size, as shown in panel (d).

The behaviour of the average density of active interfaces in the nonlinear multi-state VM is very different from that in the linear model. In the latter case the decay of $\avg{\rho}$ was found to be exponentially throughout on complete graphs, $\avg{\rho}=\avg{\rho(0)} e^{-t/\tau}$, with one single time scale $\tau=(N -1)/2$ \cite{msvm}. We re-iterate that the set of plateaux for $\rho$ differs from realization to realization in the linear model. However, the mean residence time in each plateaux and the typical `jump' in $\rho$ upon extinction of an opinion conspire such as to produce an overall exponential decay of $\avg{\rho}$ with one single time scale, independent of the number of opinions still present \cite{msvm}.  

\subsubsection{$q>1$}
\label{sec:q1}
\begin{figure}
\includegraphics[width=0.85\columnwidth]{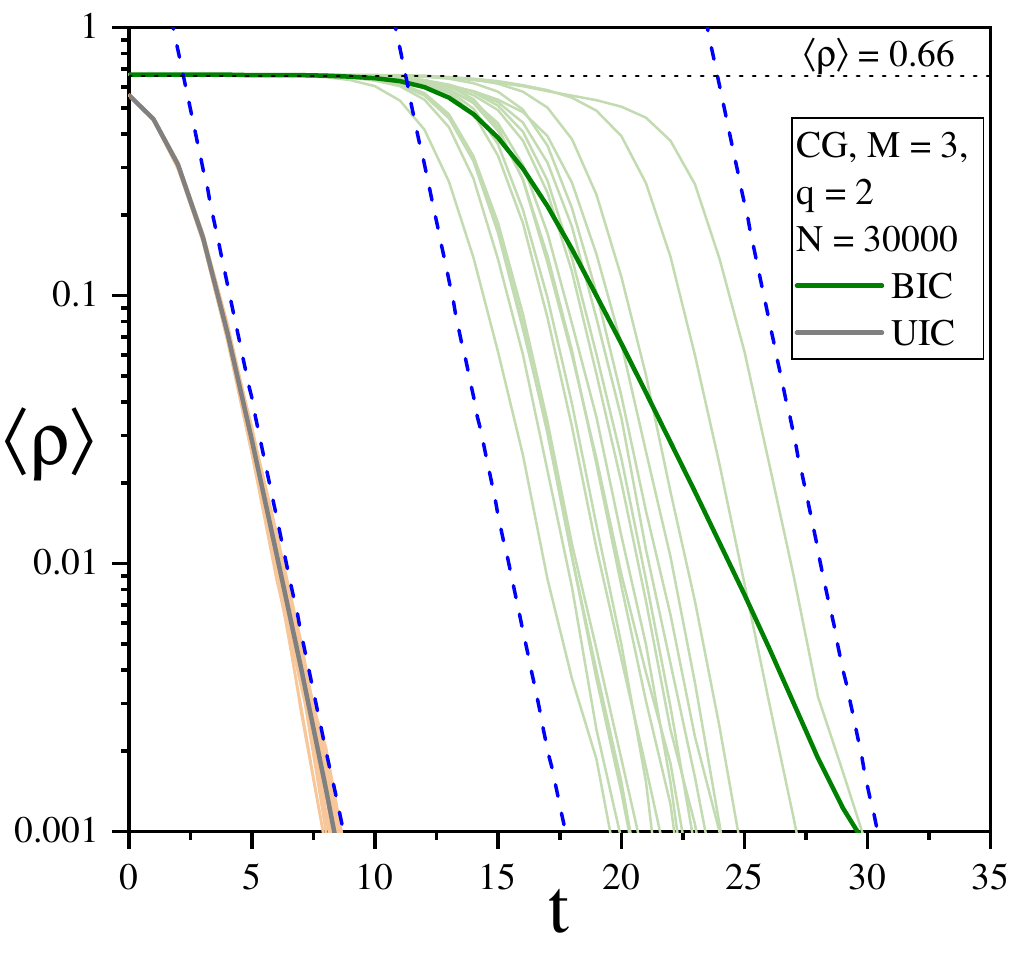}
\caption{Time evolution for the density of active links for a population of $N = 30000$ agents on a CG with  $M = 3$ and $q = 2$. Gray curve corresponds to unbalanced initial conditions (UIC)($x_1 = 3/5$, $x_2 = x_3 = 1/5$) and the dark green curve to balanced initial conditions. Individual realisations of the process are also shown; in orange for the unbalanced case  and in light green for the BIC. Blue dash lines are $\propto e^{-t}$ and are guides for the eye.\label{fig:averageq2MSVM} }   
\end{figure}

For $q>1$, the behaviour of $\rho$ is quite different from that presented above for $q<1$. For a full description, we distinguish the cases in which the initial conditions are unbalanced (UIC) or balanced (BIC). By unbalanced, we mean that when initializing the dynamics, the states of opinion are not equally represented in the population.  Both cases are illustrated in Fig.~\ref{fig:averageq2MSVM} for $M=3$ and $q=2$.  

For UIC the system approaches consensus rapidly.  The behavior of individual realisations (orange curves in Fig.~\ref{fig:averageq2MSVM}) is quite similar to the average  $\langle \rho (t)\rangle$ (gray curve).  A linear stability analysis around a stable fixed point [$(x_1,x_2,x_3)=(1,0,0)$, $(0,1,0)$, or $(0,0,1)$]  shows that in an infinite large system $\rho(t)$ approaches zero as $e^{-t}$ (see Appendix~\ref{ap:approach-M3-CG}), denoted by a dashed line, which agrees well with the exponential relaxation of $\langle \rho(t) \rangle$ from simulations. This is the same exponential relaxation as in the two-state model [Eq.~(\ref{eq:rho-t-q2})].

On the other hand, for BIC  $\avg{\rho}$ initially stays around a plateau given by $\avg{\rho}=(M-1)/M$ ($ = 0.66$ in the case of Fig.~\ref{fig:averageq2MSVM}). This is because each individual realisation is initialized in the unstable fixed point of coexistence. It then takes a time of order ${\cal O}(\log\,N)$ to escape from this fixed point via finite-size fluctuations \cite{QDT,toral_SM}. The escape time for reach realization is stochastic\cite{QDT,toral_SM}, as can be seen in the light green curves in Fig. \ref{fig:averageq2MSVM}. Once a realisation has left the area around the unstable fixed point, the decay of the interface is approximately exponential [dashed lines in Fig.~\ref{fig:averageq2MSVM} are a guide for the eye, and indicate a decay proportional to $e^{-t}$]. The initial shift, due to the random escape from the fixed point, causes the ensemble average (dark green curve) to have an exponential decay with a different decay time. 


\section{Two-state nonlinear voter model on complex networks} \label{sec:networks}

Up to now, we focused on the case of all-to-all interactions. In this section we explore the model on more complex networks. Analytical progress is here possible using approximations taking into account the degree distribution of the underlying graph \cite{sood-2005,pugliese-2009,vazquez_eguiluz,demirel-2014,vazquez-2008}. Specifically, we implement the so-called homogeneous pair approximation (HPA) developed in \cite{vazquez_eguiluz,vazquez-2008}. 

The pair approximation takes into  correlations between the states of nearest neighbours in the network, but neglects correlations between second or higher--order neighbors. In a two-state model the dynamical variables are therefore (i) the probability with which a node is in the up-state (opinion $1$), or equivalently, the magnetisation, and (ii) the probability with which any one edge is active, that is the adjacent nodes are in opposite states. The qualifier `homogeneous' indicates that these quantities are taken not to depend on the degrees of the nodes involved.  In other words, the magnetisation is the same for nodes of all degrees, and the probability to be active is the same for all edges. The HPA results in a set of two differential equations for these quantities. More sophisticated heterogeneous pair approximations have been developed to provide a more differentiated description  \cite{pugliese-2009,demirel-2014}.

The HPA is known to work well on uncorrelated networks of arbitrary degree distribution, generated by connecting pairs of nodes at random, and where degree correlations can thus be neglected \cite{vazquez_eguiluz}. Examples are degree-regular random graphs (DRRGs), Erd\"os–Renyi (ER) networks, and also Barabasi-Albert scale-free networks.  

A related type of pair approximation for the NLVM studied was implemented in \cite{min,raducha} (see Appendix~\ref{ap:minM}), where $q$ was restricted to integer numbers, and nodes are only able to change state if they are connected to $q$ or more  active links. It was also assumed that the underlying network is a DRRG, i.e., in particular, $P_k=\delta_{k,\mu}$.  This pair approximation leads to the same results as the one developed previously in \cite{jed_qvm} for the $q$--voter model. This latter model was introduced in \cite{castellano_qvm} and is similar to the NLVM, where the focal node chooses $q$ different random neighbors (without repetition) and, if they all share the same state, then the node adopts this state.

\subsection{Pair approximation}
\label{sec:hpa-M2}

In this section we develop the HPA for the two-state NLVM on complex networks for general $q>0$, following the same approach as that introduced in \cite{vazquez_eguiluz} for the linear two-state VM ($q=1$). 

Within the HPA, the time evolution of the magnetization $m = x-(1-x)$  and the density of active links $\rho$ are approximated by the  equations
\begin{subequations} 
\begin{align} 
\frac{dm}{dt}  &\approx \frac{1}{\Delta t} \sum_k \sum_{s=\pm}\sum^k_{n = 0} P(s,k,n) \nonumber\\ &~~~~~~\times P_{s \rightarrow -s }(k,n) \, \Delta m(s \rightarrow -s), 
\label{eq:m0} \\ 
\frac{d\rho}{dt}  &\approx \frac{1}{\Delta t} \sum_k \sum_{s=\pm}\sum^k_{n = 0} P(s,k,n) \nonumber\\ &~~~~~~\times P_{s \rightarrow -s }(k,n) \, \Delta \rho(s \rightarrow -s).
\label{eq:rho0} 
\end{align} 
\label{eq:dmdrdt}
\end{subequations}
Here $\Delta t=1/N$ is the elementary time step in an analogous discrete-time model. The quantity $P(s,k,n)$ is the  probability that a randomly chosen node is in state $s$ ($s=+$ or $-$), has degree $k$ and that $n$ of its neighbours are in the opposite state $-s$. Further, $P_{s \rightarrow -s }(n,k)=\left(n/k\right)^q$ is the probability that the node that was picked for update actually flips.  If the flip takes place, then the respective changes in the magnetization and the density of active links are 
 \BE
 \Delta m (s \rightarrow -s) &=& -s\frac{2}{N}, \nonumber \\
 \Delta \rho (s \rightarrow -s) &=&\frac{k-2n}{\mu N/2},
 \EE
where $\mu$ is the mean degree of the network.
 
Within the HPA, the probability $P(s,k,n)$ can be approximated as
\begin{eqnarray} 
 P(s,k,n) = x_s P_k B_s(k,n),
\end{eqnarray}
where $x_s=(1+s\,m)/2$ is the fraction of nodes in state $s=\pm 1$, $P_k$ is the degree distribution, and $B_s(k,n)$ is the probability that $n$ links connected to a node are active given that the node has degree $k$ and that it is in state $s$.  
 
Again using the HPA, $B_s(k,n)$ is a binomial distribution for $n$,
\be
B_s(k,n)=\frac{k!}{n!(k-n)!}(P_{-s|s})^n (1-P_{-s|s})^{k-n}.
\ee

Its parameters are the degree $k$ of the node, and the probability that a neighbor of a node in state $s$ is in state $-s$.
\begin{eqnarray} 
\label{eq:prom_m2}
    P_{-s|s} = \frac{\rho}{2x_s}.
\end{eqnarray}

Putting all this together we arrive at
\begin{subequations} 
\label{eq:m_rho2}
\begin{align}
\frac{dm}{dt}  &=  -2 \sum_{s=\pm} s \, x_s \sum_k    \frac{P_k }{k^q} \avg{n^q|s} , 
\label{eq:m2} \\  
\frac{d\rho}{dt}  &= \frac{2}{\mu}\sum_{s=\pm} x_s \sum_k  P_k \left( \frac{\langle n^q|s \rangle}{k^{q-1}} - \frac{2 \langle n^{q+1} |s\rangle}{k^q} \right).
\label{eq:rho2} 
\end{align} 
\end{subequations}
In these expressions $\avg{\cdots|s}$ stands for the average over the binomial distribution $B_s(k,\cdot)$, i.e., $\avg{n^q|s}$ is the $q$-th moment of the distribution. For integer values of $q$ these moments can be evaluated directly, for instance we have
\begin{eqnarray}
\label{eq:moments}
\avg{n|s} &=& k P_{-s|s},  \nonumber \\ 
\avg{n^2|s} &=& k P_{-s|s} + k(k-1)  P^2_{-s|s},  \\ 
\avg{n^3|s} &=& k P_{-s|s} + 3k(k-1)  P^2_{-s|s} \nonumber \\
&& + k(k-1)(k-2) P^3_{-s|s}. \nonumber
\end{eqnarray}
When $q$ is not an integer, the $q$-th moments can be obtained numerically for given $k$ and $P_{-s|s}$. 

For $q=1$ Eqs.~(\ref{eq:m_rho2}) become
\begin{subequations} 
\label{eq:q1}
\begin{align}
\frac{dm}{dt}  &=  0,  \\ \frac{d\rho}{dt}  &= \frac{2\rho }{\mu} \biggl\{(\mu-1)\left[1-\frac{2\rho}{(1-m^2)}\right]-1  \biggl\}.
\end{align} 
\end{subequations}
 These are the well-known results reported for example in \cite{vazquez_eguiluz}.

\subsection{Further evaluation for $q=2$}

We now briefly discuss the case $q=2$, where closed equation for $m$ and $\rho$ can be obtained. We will use these results further below. The analysis of the model with $q=2$ also highlights that network structure does not only enter through the mean degree.

For $q=2$, Eqs.~(\ref{eq:m_rho2}) become, after some algebra,
\begin{subequations} 
\label{eq:q2m2}
\begin{align}
\label{eq:dmdtq2}
\frac{dm}{dt}  &= \frac{c_1  m \rho^2}{1-m^2}, \\
\label{eq:drdtq2}
\frac{d\rho}{dt}  &= \rho \left[c_2+\frac{c_3 \rho}{1-m^2}-\frac{c_4 (1+m^2) \rho^2}{(1-m^2)^2} \right],
\end{align} 
\end{subequations}
where 
\begin{subequations}
\begin{align}
  c_1&\equiv 2(1-\mu_{-1}),\\
  c_2&\equiv 2(1-2\mu_{-1})/\mu, \\
c_3&\equiv 2(\mu-7+6\mu_{-1})/\mu, \\
  c_4&\equiv 4(\mu-3+2\mu_{-1})/\mu
\end{align}
\label{coefficients}  
\end{subequations}
are constant coefficients. In addition to the mean degree $\mu$, we have introduced $\mu_{-1}=\sum_{k=1}^\infty P_k/k$. We note that, for $m=0$, Eq.~(\ref{eq:drdtq2}) agrees with the result in \cite{peralta2018} obtained for the nonlinear noisy voter model using the HPA. While the mathematical analysis in \cite{peralta2018} describes the system for zero magnetization, Eqs.~(\ref{eq:q2m2}) also take into account the evolution of $m$ coupled to $\rho$.

Every point of the form $(m,\rho=0$) with $-1<m<1$ is a fixed point of Eqs.~(\ref{eq:q2m2}), and is linearly unstable.  The points $(1,0)$ and $(-1,0)$ are attractors of the dynamics (the system tends to evolve towards these points), and describe consensus states.  

There is a further saddle point given by 
\begin{eqnarray}
    m^* = 0, ~~~ \rho^* = \frac{c_3 + \sqrt{c_3^2 + 4c_1c_4}}{2c_4},
    \label{m-rho-fixed}
\end{eqnarray}
which is linearly unstable in the $\vec{v}_1=(1,0)$ direction and linearly stable in the $\vec{v}_2=(0,1)$ direction, with associated eigenvalues $\lambda_1=c_2 (\rho^*)^2>0$ and $\lambda_2=-\rho^* \sqrt{c_3^2+4c_2c_4}<0$. This represents an active state ($\rho^*\neq 0$) in which half of the nodes are in either state $\pm 1 $.

\subsection{Dynamical behaviour and comparison against simulations}
\label{sec:sim-networks}

\begin{figure}
\centering
{\includegraphics[width=0.8\columnwidth]{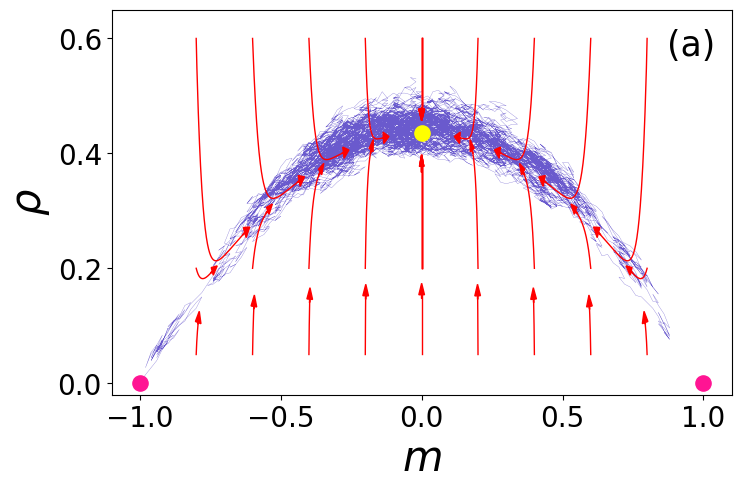}} \\
{\includegraphics[width=0.8\columnwidth]{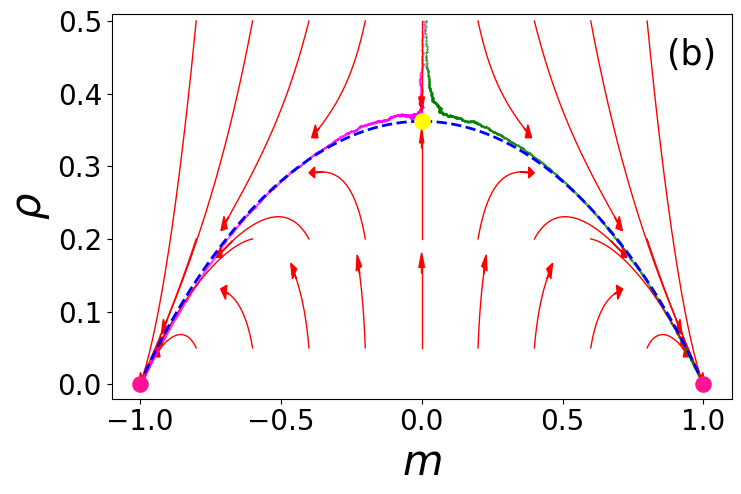}}
\caption{Trajectories of the system in a single realisation on the $m$--$\rho$ space starting from a point near the symmetric initial condition $(0,0.5)$, for an ER network of mean degree $\mu=8$, $N=100$ nodes and $q=0.9$ (a), and $N=10^4$ and $q=2$ (b).  The two pink filled circles denote the attactors $(-1,0)$ and $(1,0)$, which are unstable for $q=0.9$ (a) and stable for $q=2$ (b).  The yellow filled circle denotes the stable fixed point $(0,0.435)$ in panel (a), and the unstable fixed point $(0,0.362)$ in panel (b).  Solid lines correspond to the deterministic trajectory obtained from the numerical integration of Eqs.~(\ref{eq:m_rho2}) for panel (a) and Eqs.~(\ref{eq:q2m2}) for panel (b), where arrowheads indicate the flow's direction.  The blue dashed line in (b) corresponds to the deterministic trajectory starting from two different almost symmetric initial conditions, $(-0.001,0.362)$ and $(0.001,0.362)$.  }
\label{fig:m-rho}
\end{figure}

Fig.~\ref{fig:m-rho} illustrates trajectories obtained from Eqs.~(\ref{eq:m_rho2}) in the $(m,\rho)$ space for different values of $q$. As seen in the figure, these are in agreement with trajectories from simulations in finite populations.

\begin{figure}
\centering
\includegraphics[width=0.8\columnwidth]{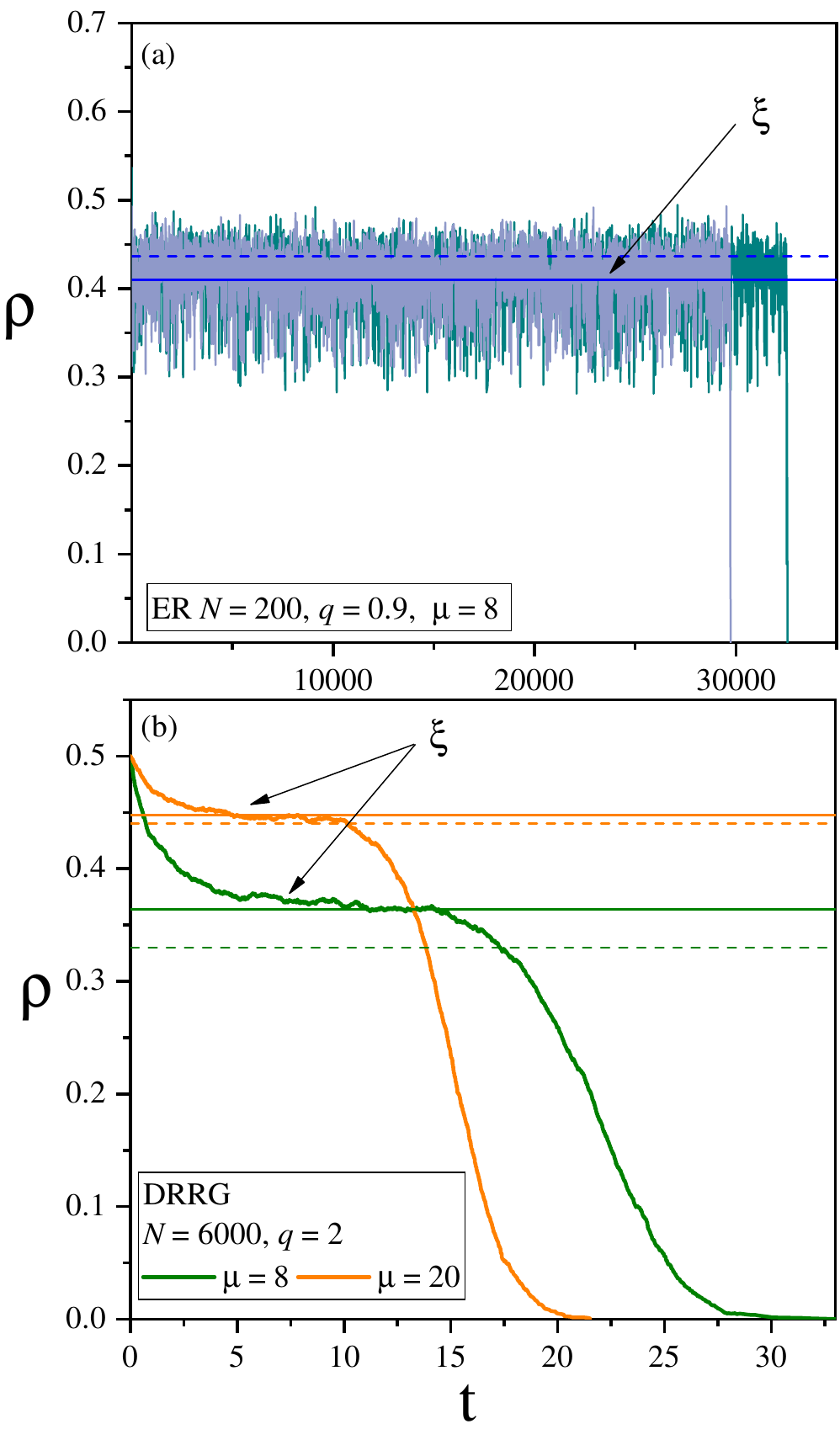}
\caption{Time evolution of the density of active links for single realisations of the binary NLVM. Panel (a) shows for $q = 0.9$ on ER of $N=36$ with $\mu=8$, and panel (b) for $q = 2$, on DRRGs of $N=6000$ and mean degrees $\mu=8$ and $\mu=20$.  When $q<1$ realizations stays around a $\rho$ value we call $\xi$ until a finite size transition takes the system to consensus. For $q>1$ we see that $\rho$ rapidly decays to a plateau value $\xi$, where it remains for a short time until it eventually decays to zero.  Solid horizontal lines are the predictions of the HPA [Eq.~(\ref{eq:m_rho2}) (Eq.~(\ref{m-rho-fixed}) for (b))]. Dashed lines are LDPA-prediction in Eq.~(\ref{eq:Min-rho-M=2}).}
\label{fig:m2rr}
\end{figure}

We run simulations on a finite network, where each node is assigned states $\pm 1$ randomly (with equal probability) and independently. In Fig.~\ref{fig:m2rr} we plot the time evolution of the density of active links $\rho$ for single realisations on Erd\"os-Renyi graphs. $\rho$ quickly evolves towards a quasi-stationary value broadly consistent with the value of $\rho$ at the unstable fixed point in Eq.~(\ref{eq:m_rho2}). This corresponds to the apex of the concave curve (inverse U-shape) in the $(m,\rho)$ space shown in Fig.~\ref{fig:m-rho}. In Fig.~\ref{fig:m2rr} we also compare the plateau value for $\rho$ with that predicted in \cite{min, raducha} (see also Appendix~\ref{ap:minM})
\be
\rho^*_{\rm LD}(M=2)=\frac{\mu-2q}{2(\mu-q)},
 \label{eq:Min-rho-M=2}
\ee
shown in Fig.~\ref{fig:m2rr} as dashed lines.  As can be seen in Fig.~\ref{fig:m2rr}(b), $\rho^*_{\rm LD}$ becomes more consistent with simulations for bigger values of $\mu$. Thus we refer to this as the as the `Large Degree Pair Approximation' (LDPA), as also indicated by the subscript `LD'. The approximation in Eq.~(\ref{m-rho-fixed}) provides a good estimate, while the approximation from \cite{min,min2,raducha} underestimates simulation results, in particular for small mean degree $\mu$.

The difference between the results given by the  HPA  (Eq.~(\ref{eq:m_rho2})) and LDPA (Eq.~(\ref{eq:Min-rho-M=2})) lies in how the rate equations are constructed. Although the update rule in both cases is the same,  the LDPA considers an approximation that implies that to make a flip, exactly $q$ neighbors are selected without repetition; that approximation facilitates calculations and gives consistent values for large $\mu$ but no so for low $\mu$  (the LDPA is further described in Appendix~\ref{ap:minM}). Fig.~\ref{fig:m2q2} provides further comparison of the performance of both methods against results form simulations for different $q$ values and a broad range of $\mu$. It should be noted though that the approximation of Ref.~\cite{min,min2,raducha} has the advantage of giving a simple expression for general $q$. For large values of $\mu$ both expressions tend to $\rho = 1/2 $, which corresponds to the fixed point on a CG [Eqs.~(\ref{eq:m_rho2})].

\begin{figure}
\centering
\includegraphics[width=0.85\columnwidth]{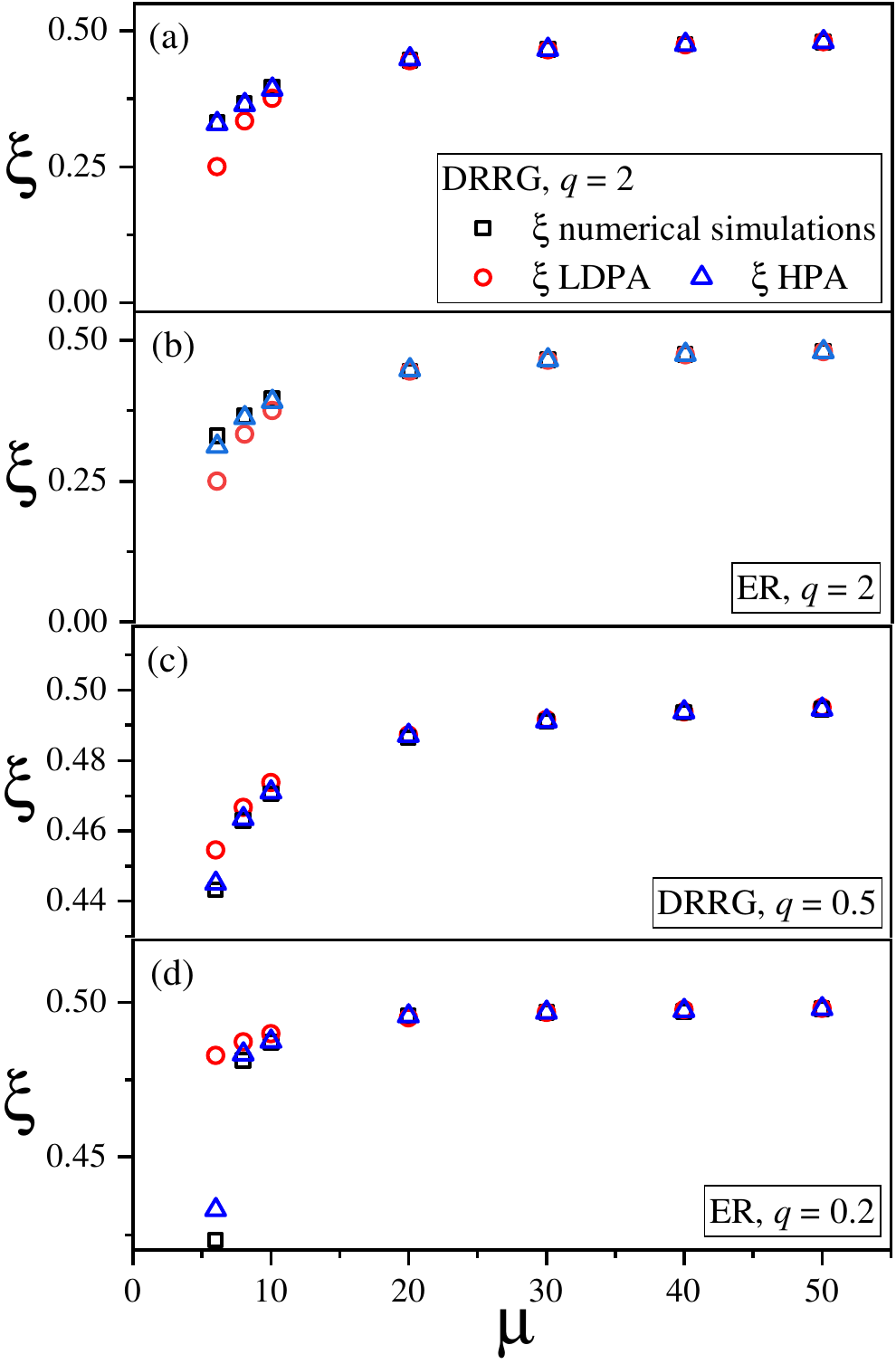}
\caption{Plateau value $\xi$ vs. mean degree $\mu$ in the binary NLVM on networks. Balanced initial conditions are considered. Squares correspond to simulations data, circles to the LPDA, and triangles to the HPA. In panels (a) and (b) $q=2$ on a DRRG and an ER network, respectively; the value of $\xi$ for the HPA is that from Eq.~(\ref{m-rho-fixed}).  Panels (c) and (d) correspond to a DRRG with $q=0.5$ and an ER network with $q=0.2$, respectively.  The value of $\xi$ for the HPA corresponds to the numerical integration of Eqs.~(\ref{eq:m_rho2}).  While the HPA gives accurate values of $\xi$ for $\mu \gtrsim 6$, the approximation from \cite{min} works well for $\mu \gtrsim 20$.}
\label{fig:m2q2}
\end{figure}

\subsection{Final approach to consensus}

The approach to consensus on complex networks is qualitatively similar to that on CG studied in Secs.~\ref{superlinear} and \ref{sublinear}, which has two distinct behaviors depending on the value of $q$.  As we described in Sec.~\ref{sec:sim-networks}, for $q < 1$ the system evolves towards the stable fixed point $(0,\rho^*)$ and eventually reaches consensus by finite-size fluctuations, in a time that grows exponentially with $N$.  Therefore, the average density of active links decays to zero with a rate given by the inverse of the mean consensus time $\tau$, i.e., $\langle \rho(t) \rangle\approx \rho^* e^{-t/\tau}$.  For $q > 1$ there is a fast (logarithmic in $N$) approach to consensus as a consequence of the drift towards one of the two absorbing states. The mean consensus time also depends on the topology of the network and the initial condition.  

To gain further insight into the relaxation towards consensus for $q>1$, we study the case $q=2$ within the HPA developed in Sec.~\ref{sec:hpa-M2} [in particular Eqs.~(\ref{eq:q2m2})].  As these equations are non-linear, an analytical exact solution is hard to obtain.  Still, valuable information can be extracted from Eqs.~(\ref{eq:q2m2}) by investigating how the system relaxes to the stable consensus fixed points $(1,0)$ and $(-1,0)$. 

In Appendix~\ref{ap:approach} we show that for an arbitrary initial condition $m(0)>0$ and $\rho(0)>0$ the asymptotic approach to $(1,0)$ is of the form $m(t)=1-\epsilon(t)$ and $\rho(t)= \alpha \epsilon(t)$, where $\epsilon(t)$ is a function that satisfies $0 < \epsilon(t) \ll 1$ and
\be
  \alpha = \frac{c_3 + \sqrt{c_3^2+8(c_4-c_1)c_2}}{2(c_4-c_1)}
  \label{alpha}
\ee
is a constant.  Then, in the limit of large times, Eqs.~(\ref{eq:q2m2}) become 
\be
    \frac{d\epsilon(t)}{dt} = - \frac{c_1 \alpha^2}{2} \epsilon(t)
\ee
to first (non-zero) order in $\epsilon$. One then has $\epsilon(t) = \epsilon_0 \, e^{-c_1 \alpha^2 t/2}$.  Thus, the relaxation towards the attractor $(1,0)$ is given by the expressions

\begin{eqnarray}  
  \label{m-t-ap}
  m(t) &=& 1-(1-m_0)e^{-c_1 \alpha^2 t/2}, \\
  \label{rho-t-ap}
  \rho(t) &=& \alpha (1-m_0)e^{-c_1 \alpha^2 t/2}. 
\end{eqnarray}

Estimating the mean consensus time $\tau$ from the condition $m(\tau)=1-2/N$, we obtain from Eq.~(\ref{m-t-ap})
\begin{eqnarray}
  \tau \simeq \frac{2}{c_1 \alpha^2} \ln \left[ \frac{1}{2}(1-m_0)N \right].
  \label{tau-N}
\end{eqnarray}
This estimate depends on the network's topology through the constants $\alpha$ and $c_1$, which in turn contain the mean degree $\mu$ and the quantity $\mu_{-1}$ [see Eqs.~(\ref{coefficients}) and (\ref{alpha})].  In the high-connectivity limit $\mu\gg 1$, the expression in Eq.~(\ref{tau-N}) agrees with that for complete graphs $\tau \simeq \ln \left[ \frac{(1-m_0^2)N}{4 m_0^2} \right]$ for $m_0 \lesssim 1$, obtained from Eq.~(\ref{eq:tau-N-q2}) by using $\rho_0=(1-m_0^2)/2$.

This analysis suggests that the scaling of the mean consensus time with the population size is of the form $\tau \simeq a \ln N + b$, where $a$ and $b$ depend on the network structure through the moments $\mu$ and $\mu_{-1}$ and, additionally, $b$ carries a dependence on the initial magnetisation $m_0$ [see Eq.~(\ref{tau-N})].  

\begin{figure}
\centering
\includegraphics[width=1\columnwidth]{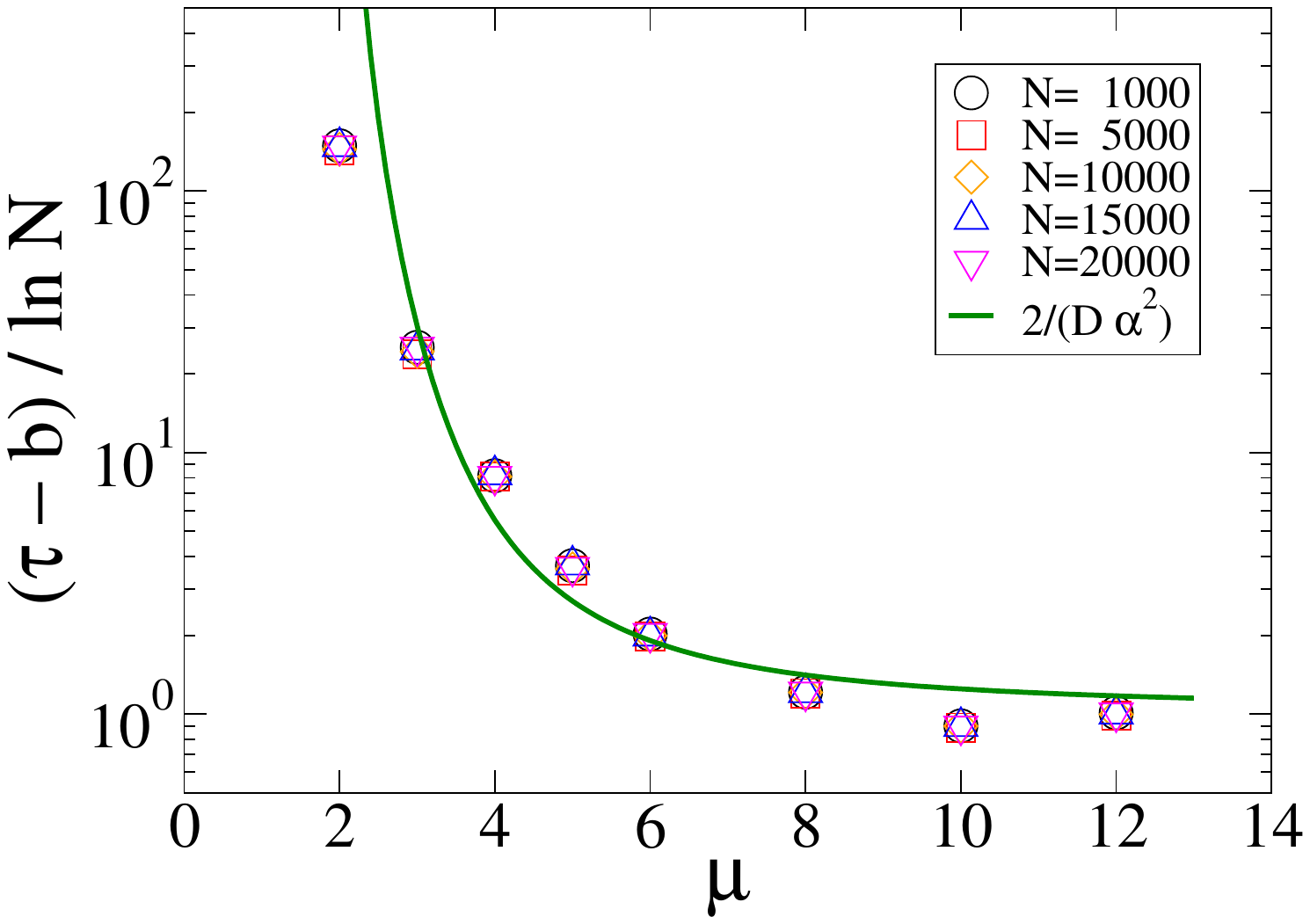}
\caption{Scaled mean consensus time $\tau$ vs mean degree $\mu$ for the $q=2$ NLVM on ER networks of various sizes $N$, as indicated in the legend.  The factor $b$ is the $y$-intercept obtained by a linear fit of the curves $\tau$ vs $\ln N$.  The solid line is the analytical prediction $2/(c_1 \alpha^2)$ from Eq.~(\ref{tau-N}).} 
\label{fig:tau-b-mu}
\end{figure}

Test against simulations are presented in Fig.~\ref{fig:tau-b-mu}. We observe a good collapse of the data for the different network sizes, and that the approximation $2/(c_1 \alpha^2)$ appears to be reasonably accurate for $\mu \gtrsim 3$, while deviations are seen for lower mean degree. A similar overestimation of the mean consensus time calculated with the HPA for low degrees was found in the linear VM on ER networks \cite{vazquez_eguiluz}, where the estimated value of $\tau$ diverges as the mean degree approaches the value $\mu \simeq 2$.  This is because the HPA assumes that the macroscopic quantities $x_{\alpha}$ and $\rho$ are independent on the nodes' degree, which is not longer valid for networks with low connectivity.  Indeed, a very accurate analytical expression was given by means of an heterogeneous pair approximation \cite{pugliese-2009}.

\section{Pair Approximation for the multistate NLVM on complex networks}\label{sec:pa_qmsvm}

We now consider the multistate NLVM on networks. We focus here on determining the different plateaux values of the density of active links and their dependence on the network degree distribution. The behavior of the ensemble  average of $\rho$ is qualitatively the same than for the Complete Graph (Section IV), with the plateaux values of that case replaced by the ones we determine in this section. We begin by a summary of the behaviour  seen in simulations, and then move on to develop the pair approximation approach.

\subsection{Numerical simulations} \label{sec:numsimNetworks}

\begin{figure}
\centering
\includegraphics[width=0.80\columnwidth]{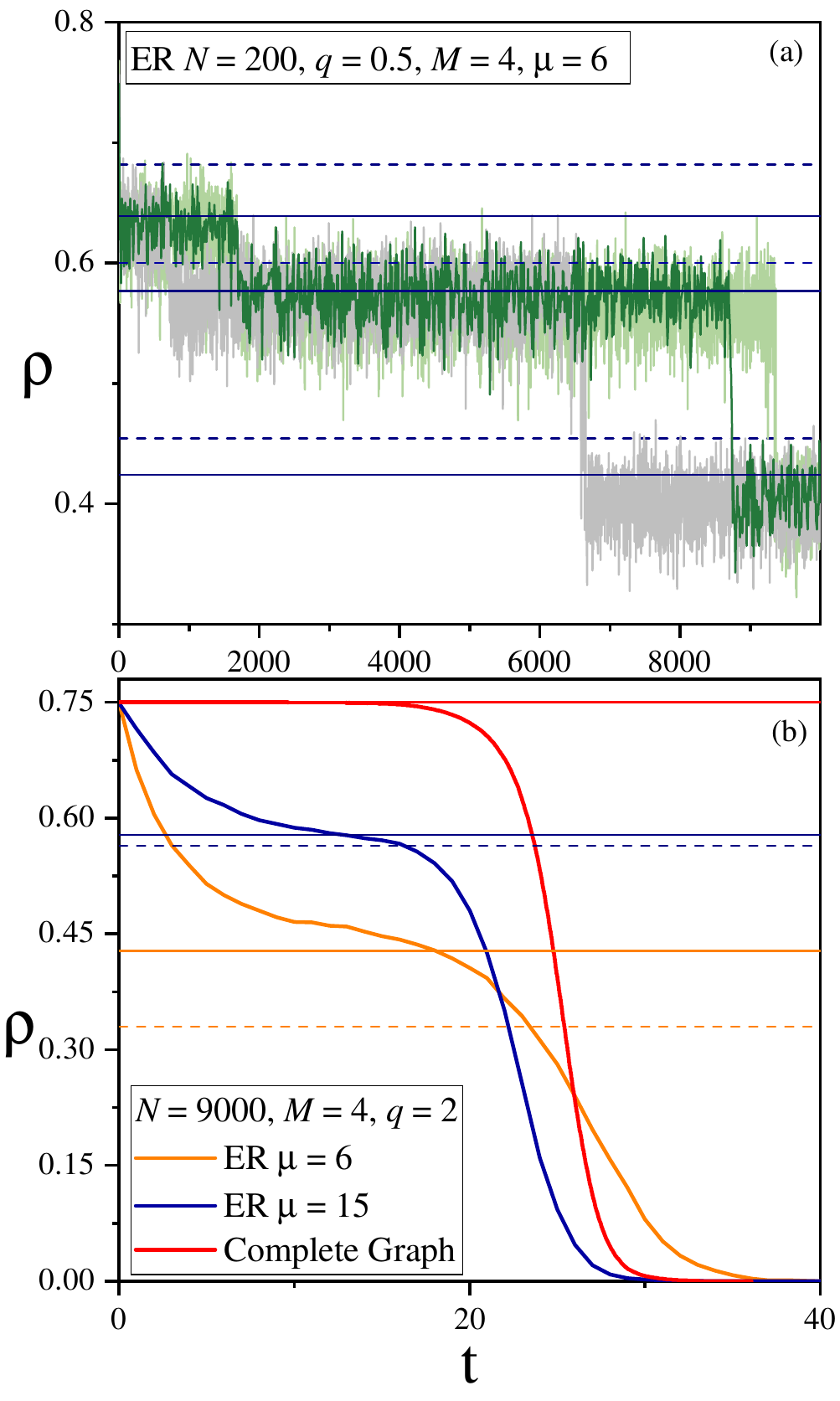}
\caption{Time evolution of the density of active links $\rho$ in individual realisations of the multistate NLVM for $M=4$ opinion states on ER graphs starting from balanced initial conditions.  Panel (a) is for $q=0.9$, panel (b) for $q=2$. In (b) we also show results for complete graphs for comparison CG. Solid horizontal lines are the predictions of the HPA [Eq.~(\ref{eq:fp_msvm})]. Dashed lines are LDPA-prediction in Eq.~(\ref{eq:min_generalM}).
}
\label{fig:ir_er_m4}
\end{figure}

Fig.~\ref{fig:ir_er_m4} shows the time evolution of the density of active links $\rho$ in single realisations of the NLVM on ER networks  for $M=4$ opinions and two values of $q$. We observe the presence of intermediate plateaux when $q<1$ corresponding to $4$, $3$, and $2$ surviving opinions [panel (a)]. These are not present for $q>1$ [panel (b)]. We thus conclude that individual realisations of the NLVM with multiple opinion states on networks behave qualitatively similarly to those in the model on complete graphs (compare with Fig.~\ref{fig:1realisations_cg}).

The plateau values for the density of active interfaces can be estimated analytically through 
an extension of the pair approximation for the model with two opinions \cite{min}. This is described in more detail in Appendix~\ref{ap:minM}.  In this approximation we obtain explicit analytical expressions for the stationary value of $\rho$ when $M$ opinions survive that generalizes Eq.~(\ref{eq:Min-rho-M=2}): 
\be \label{eq:min_generalM}
\rho^*_{\mbox{\tiny LD}} = \frac{(M-1) (\mu-2q)}{M(\mu-q)},
\ee

For $q=1$, Eq.~(\ref{eq:min_generalM}) reduces to the accurate expression for the plateau values obtained in \cite{msvm} for balanced initial conditions.  However, the approximation leading to Eq.~(\ref{eq:min_generalM}) has limitations for $q \ne 1$. For example, it is not accurate for small values of $\mu$, see in Fig.~\ref{fig:xi_vs_mu}.  Moreover, if $q<\mu<2q$ one obtains a non-physical value $\rho^*_{\mbox{\tiny LD}}<0$.  In order to obtain a better approximation we next extend the HPA developed in Sec.~\ref{sec:networks} to the general case of $M$ opinions.

\subsection{Homogeneous pair approximation for the general case $M \ge 2$}

\subsubsection{General formalism}

We now construct the homogeneous pair approximation for the model with multiple opinion states. More specifically, our aim is to derive dynamical equations analogous to those for the two-state NLVM [Eqs.~(\ref{eq:dmdrdt})]. We follow an approach similar to that in Sec.~\ref{sec:hpa-M2}.

Suppose a node in state $\alpha$ and with degree $k$ is chosen for potential update.  We denote by $n_\beta$ the number of neighbours in state $\beta$.  A flip $\alpha \rightarrow \beta$ ($\beta \ne \alpha$) of the focal node occurs with probability $(n_{\beta}/k)^q$.  Then, the number of active links before and after the update is $k-n_{\alpha}$ and $k-n_{\beta}$, respectively, and thus the change in the density of active links $\rho$ is $(n_{\alpha}-n_{\beta})/(\mu N/2)$.  Summing over all possible node updates we arrive at the following rate equation for $\rho$: 
\begin{eqnarray}
    \label{eq:aux1}
    \frac{d\rho}{dt} &=& \frac{1}{1/N} \sum_k \sum^M_{\alpha = 1} \sum^M_{\beta \neq \alpha} x_{\alpha} P_k \times \\
    && \sum_{\bn} {\cal M}_\alpha(k,\textbf{n}) \left(\frac{n_\beta}{k}\right)^q \frac{(n_{\alpha} - n_{\beta})}{\mu N/2}. \nonumber
\end{eqnarray}

Here ${\cal M}_\alpha(k,\textbf{n})$  represents the probability that a node of degree $k$ and state $\alpha$ has $n_1$ neighbours in state $1$, $n_2$ neighbours in state $2$ and so on. We have written $\textbf{n} = (n_1,n_2, \dots, n_M$). 

Analogous to the two-state NLVM, we neglect state correlations between second or higher-order neighbours, and approximate $\mathcal M_{\alpha}(k,\textbf{n})$ as a multinomial distribution 
\BE\label{eq:multinomial}
{\cal M}_\alpha(k,\textbf{n}) &=& \frac{k!}{n_1! n_2!\cdots n_M!}\nonumber \\
&& \hspace{-2em} \times (P_{1|\alpha})^{n_1} (P_{2|\alpha})^{n_2}\cdots (P_{M|\alpha})^{n_M},
\EE
 where $P_{\beta|\alpha}$ is the probability that a randomly chosen neighbour of a node in state $\alpha$ is in state $\beta$. Within the pair approximation and similar to Eq.~(\ref{eq:prom_m2}), this probability is given by
\be \label{eq:prom_m}
 P_{\beta|\alpha} =\frac{\rho_{\beta\alpha}}{2x_\alpha},
\ee
where $\rho_{\beta\alpha}$ is the fraction of links connecting nodes in states $\alpha$ and $\beta$.

We now focus on the stationary state, setting $d\rho/dt=0$ in Eq.~(\ref{eq:aux1}). Since all opinion states are equivalent all opinions are equally represented in the stationary state by symmetry, i.e., $x_{\alpha}=1/M$ for $\alpha=1,\dots,M$. Given that all $M(M-1)$ terms contribute equally to the sum in Eq.~(\ref{eq:aux1}) we arrive at  
 \begin{align}\label{eq:aux2}
 \frac{2(M-1)}{\mu}\sum_k x_{\alpha} P_k
\sum_{\bn} {\cal M}_\alpha(k,\textbf{n}) \left(\frac{n_\beta}{k}\right)^q (n_{\alpha} - n_{\beta}) &= 0
\end{align}
for any particular combination of $\alpha$ and $\beta$. From this we find
\begin{equation} \label{eq:fp_msvm}
\sum_k x_{\alpha} \frac{P_k}{k^q} \left( \avg{n_{\alpha} n_{\beta}^q}_{\cal M_{\alpha}} -\avg{n_{\beta}^{q+1}}_{\cal M_{\alpha}}  \right) =0,
\end{equation}
where $\avg{\dots}_{\cal M_{\alpha}}$ stands for averages over the multinomial distribution in Eq.~(\ref{eq:multinomial}). We note that $\mathcal M_{\alpha}$ carries a dependence on $k$.

\subsubsection{Linear model ($q=1$)}

When $q = 1$ Eq.~(\ref{eq:fp_msvm}) becomes
\be \label{eq:fp_msvm_q1}
\sum_k x_{\alpha} \frac{P_k}{k} \left( \avg{n_{\alpha}n_{\beta}}_{\cal M_{\alpha}} -\avg{n_{\beta}^2}_{\cal M_{\alpha}}\right)=0.
\ee
The relevant moments of $\mathcal M_{\alpha}$ are 
\BE \label{eq:moments_m}
 \avg{n_{\alpha}n_{\beta}}_{\cal M_{\alpha}} &=& k (k-1)  P_{\alpha|\alpha}  P_{\beta|\alpha} \nonumber \\ 
\avg{n_\beta^2}_{\cal M_{\alpha}} &=& k P_{\beta|\alpha} + k(k-1) P_{\beta|\alpha}^2.
\EE

Using Eqs.~(\ref{eq:moments_m}) and the fact that $P_{\alpha|\alpha}=1-\sum_{\gamma\neq\alpha} P_{\gamma|\alpha}$ in Eq.~(\ref{eq:fp_msvm_q1}), we arrive at
\BE \label{eq:fp_msvm_q11}
&&\sum_k x_\alpha P_k P_{\beta|\alpha}\bigg\{   (k-1)(1-\sum^M_{m\neq \alpha} P_{m|\alpha})-1 \nonumber \\
&&-(k-1)P_{\beta|\alpha}\bigg\}=0.
\EE
The symmetry of opinion states implies that $\rho_{\alpha\beta}$ takes the same value for all $M(M-1)$ combinations $\alpha \neq \beta$. Noting $\rho=\sum_{\alpha\neq\beta} \rho_{\alpha\beta}/2$, we then have $\rho_{\beta\alpha}=2\rho / [M(M-1)]$ in Eq.~(\ref{eq:prom_m}). Using this, and $x_\alpha=1/M$, we have
\be
P_{\beta|\alpha} = \frac{\rho}{M-1},
\ee
for $\alpha\neq \beta$. The probability that a link is inactive is $P_{\alpha|\alpha} = 1-\sum^M_{m\neq \alpha} P_{m|\alpha} = 1-\rho$. Putting everything together, and carrying out the remaining sum over $k$ in Eq.~(\ref{eq:fp_msvm_q1}) we arrive at
\be
\label{eq:fp_msvm_linear}
\frac{2}{\mu}  [(\mu-2)-\frac{M}{M-1}(\mu-1)\rho ]=0, 
\ee
and therefore at the fixed-point solution
\be
\rho^* = \left( 1 - \frac{1}{M} \right)\frac{\mu-2}{\mu-1}.
\ee
This reproduces the result of \cite{msvm}, where the linear multistate VM was first reduced to an effective two-state model, and where established pair-approximation results for two-states models \cite{vazquez_eguiluz} were then exploited.

\subsubsection{Further evaluation of Eq.~(\ref{eq:fp_msvm})  and test in simulations}
When $q = 2$ Eq.~(\ref{eq:fp_msvm}) becomes 
\be \label{eq:fp_msvm_q2}
\sum_k x_{\alpha} \frac{P_k}{k^2} \left( \avg{n_{\alpha}n_{\beta}^2}_{\cal M} -\avg{n_{\beta}^3}_{\cal M} \right)=0,
\ee
and we have
\begin{subequations} 
\label{eq:moments_q2}
\begin{align*} \avg{n_{\alpha}n_{\beta}^2}_{\cal M} &= k (k-1)  P_{\alpha|\alpha}  P_{\beta|\alpha} \\ &+ k(k-1)(k-2) P_{\alpha|\alpha}  P_{\beta|\alpha} ^2\\ 
\avg{n_\beta^3}_{\cal M} &= k P_{\beta|\alpha} + k(k-1) P_{\beta|\alpha}^2  \\
&+ k(k-1)(k-2)  P_{\beta|\alpha} ^3 . 
\end{align*} 
\end{subequations}
Using this in Eq.~(\ref{eq:fp_msvm_q2}) and after some algebra, we find the following plateau density for the model with $q=2$ and with $M$ opinion states,
\be \label{eq:rho*q2}
\rho^*_{q=2,M} = \frac{-B \pm \sqrt{B^2 -4AC}}{2A}
\ee
where
\begin{subequations} \label{eq:q2}
\begin{align} A &= - \frac{M}{(M-1)^2}(\mu+2\mu_{-1}-3) \\ B &= \frac{1}{M-1}[(\mu+2\mu_{-1}-3)-(M+2)(1-\mu_{-1})] \\ C&=1-2\mu_{-1}, \end{align} \end{subequations}
We recall that the quantity $\mu_{-1}$ is the mean inverse degree, $\sum_k P_k/k$.  

\begin{figure}
\centering
\subfloat{\includegraphics[width=0.85\columnwidth]{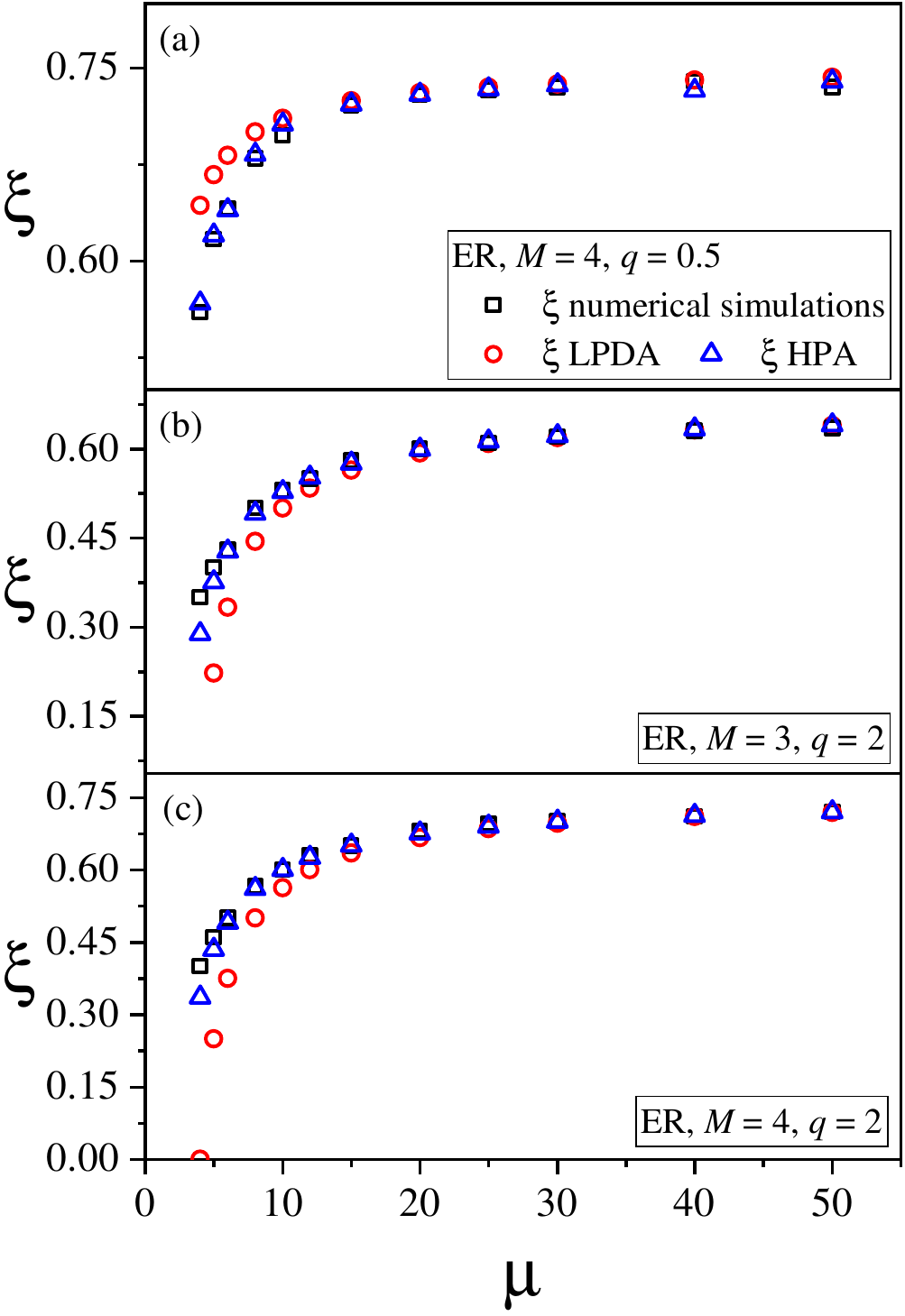}}
\caption{Plateau value $\xi$ for the mean density of active interfaces vs. mean degree $\mu$ in the model on ER networks. For $q<1$ [panel (a)] we show data for the first plateau in the time-evolution of the system (i.e., all $M$ opinions are present). For $q>2$ we use balanced initial conditions, and show the interface density during the initial escape from the state with all $M$ opinions. Simulations data is shown as squares. Triangles show the predictions of the HPA [Eq.~(\ref{eq:fp_msvm}) for panel (a), Eq.~(\ref{eq:rho*q2}) for (b) and (c)]. Predictions from the  LDPA [Eq.~(\ref{eq:min_generalM})] are shown as circles.}\label{fig:xi_vs_mu}
\end{figure}

\subsection{Test against simulations}
 We now test the predictions of the HPA against simulation results. The main quantity we assess is the density of active interfaces at the different plateaux during the time-evolution of the system. For $q<1$ these plateaux emerge from the sequence of extinctions [as illustrated in Fig.~\ref{fig:ir_er_m4}(a)]. For $q>1$ we focus on the initial plateau after a start from balance initial conditions [Fig.~\ref{fig:ir_er_m4}(b)].

Results can be found in Fig.~\ref{fig:xi_vs_mu}. We find that the predictions of the HPA agree better with simulations than those of the large-degree pair approximation [Eq.~(\ref{eq:min_generalM})], especially for small values of the mean degree. This can also be seen  in Fig.~\ref{fig:ir_er_m4}, where solid lines (HPA) are closer to simulation results that dashed lines (LDPA).

 We highlight that the HPA involves the network structure not only through the mean degree. For example, when $q = 2$ the mean inverse degree $\mu_{-1}$ enters into the prediction [see Eqs.~(\ref{eq:rho*q2}), (\ref{eq:q2})]. For general powers $q$ of the nonlinearity further moments of the degree distribution will be required.

\section{Summary and conclusions}
\label{sec:conclusions}

We have studied the ordering dynamics in voter models with nonlinear imitation rules. More precisely, we have taken the rate with which an individual adopts an opinion to scale as the $q$-th power of the number of its neighbours in this state. We here summarise our main results, before we draw more general conclusions and discuss the wider implications. The main features of the different models and of the ordering dynamics in them are also collated in Tables \ref{table:cg} and \ref{table:graphs}.

\begin{table*}
\centering
\begin{tabular}{|c|c|c|c|}
\toprule
~~~{\bf CG}~~~& {\bf drift} & {\bf plateaux for $\rho$ in} & {\bf ordering/behaviour of}  \\  
 ~&   & {\bf individual realizations} & {\bf ensemble average}  \\
\hline
$q<1$ &towards & each realization has $M-1$ plateaux,  & escape from metastable states by fluctuations  \\
        &  coexistence  & located at $\xi_\ell=(\ell-1)/\ell$ & $\avg{\rho}-\avg{\rho}_\ell \sim  e^{-t/\tau_\ell}$ \\
        & &($\ell=2,\dots, M$) & $\tau_\ell\sim e^{f(\ell, q)N}$ \\
        && [Figs.~\ref{fig:1realisations_cg}(a), \ref{fig:rho_qsmaller1_cg}]& [Figs.~\ref{fig:decayMSVM}, \ref{fig:scalingMSVM}] \\[0.5em] \hline
$q=1$ & none & each realization has $M-1$ plateaux, & driven by fluctuations \\
&  &  the first plateau is at $(M-1)/M$ & $\avg{\rho}=\rho_0 e^{-t/\tau}$ \\
&  & and the rest can be at any value & $\tau\sim N$\\
&  & [Fig.~\ref{fig:1realisations_cg}(b)], Ref.~\cite{msvm}  &  Ref.~\cite{msvm}\\[0.5em] \hline
$q>1$ & towards  & none & follows deterministic rate equations\\
& consensus& [except for noise-driven escape from balanced initial & $\avg{\rho}\sim e^{-t}$ ($t\gg 1$)\\
& &conditions with a plateau at $\rho=(M-1)/M$)]& mean consensus time $\tau\sim \ln\,N$ \\
& & [Fig.~\ref{fig:1realisations_cg}(c)] &[Fig.~\ref{fig:averageq2MSVM}]  \\
\hline\hline
\end{tabular}
\caption{Overview of behaviour of individual realizations and ordering dynamics for nonlinear multi-state VMs on complete interaction graphs. \label{table:cg}}
\end{table*}

\begin{table*}
\centering
\begin{tabular}{|c|c|c|c|}
\toprule
{\bf uncor-}& {\bf drift} & {\bf plateaux for $\rho$ in} & {\bf ordering/behaviour of }  \\  
{\bf -related}&   & {\bf individual realizations} & {\bf ensemble average}  \\ 
{\bf graphs}&   &   &   \\ 
\hline
$q<1$ &towards & each realization has $M-1$ plateaux  & escape from metastable states by fluctuations  \\
        &  coexistence  & located at $\xi_\ell(\{P_k\},q)$& $\avg{\rho}-\avg{\rho}_\ell \sim  e^{-t/\tau_\ell}$  \\
        & &($\ell=2,\dots, M$) &$\tau_\ell\sim e^{f(\ell,\{P_k\}, q)N}$ \\
        &&[Figs.~\ref{fig:ir_er_m4}(a), \ref{fig:xi_vs_mu}]&\\[0.5em] \hline
$q=1$ & none & each realization has $M-1$ plateaux & by fluctuations \\
&  &  first plateau at $\xi_M(\mu)$,  & $\avg{\rho}=\xi_M(\mu) e^{-t/\tau}$ \\
&& any value of $\rho$ possible &$\tau\approx {\cal A} N$ ($N\gg 1$), ${\cal A}={\cal A}(\{P_k\})$\\
&  &  for subsequent plateaux  &  Ref.~\cite{msvm}  \\
& &Ref.~\cite{msvm} & \\[0.5em] \hline
$q>1$ & towards  & one initial plateau if initial conditions are balanced   & driven by deterministic pull \\
& consensus& [Fig.~\ref{fig:m2rr}(b),~Fig.~\ref{fig:ir_er_m4}(b)] & $\avg{\rho}\sim e^{-t/{\cal B}}$ ($t\gg 1$) \\
&&& consensus time $\tau\approx {\cal B}\ln\,N$ ($N\gg1$),  ${\cal B}={\cal B}(\{P_k\})$  \\
& & & [Eqs.~(\ref{rho-t-ap}, \ref{tau-N}) demonstrate this for $M=2, q=2$] \\
\hline\hline
\end{tabular}
\caption{Overview of ordering dynamics for nonlinear multi-state VMs on uncorrelated interaction graphs, with $N$ nodes and degree distribution $P_k$. In general the coefficients ${\cal A}$ and ${\cal B}$ depend on the entire degree distribution, $\{P_k\}$. Within the HPA the coefficient ${\cal A}$ only depends on the first and second moments of $\{P_k\}$ \cite{msvm}. For $q=2$ and within the HPA the coefficient ${\cal B}$ only depends on $\mu$ and $\mu_{-1}$. The same is true for any $q>1$ in the LDPA. We confirmed the ordering and behaviour of the average density of active links for $q<1$ in simulations, but this data is not explicitly shown in the paper.\label{table:graphs}}
\end{table*}

\subsection{Summary of main results}
\subsubsection{No reduction to a two-state model when the dynamics are nonlinear}
It is well-known that, for many purposes, it is possible to reduce the dynamics of linear multi-state voter models ($q=1$) to effective two-state systems \cite{redner, herrerias, msvm}. This is because the dynamics of the number of agents in one focal opinion does not depend on the detailed numbers of agents in each of the other opinion states. No such reduction can be made in nonlinear voter models, complicating the quantitative analysis. Nonetheless, we have been able to characterise the ordering process in the nonlinear model using a combination of simulations and analytical methods.

\subsubsection{Ordering dynamics is noise-driven for $q<1$, but not for $q>1$}
Depending on the value of $q$, we have find two distinct types of ordering dynamics. This distinction applies both in models with all-to-all interaction and on graphs. 

The first type of dynamics is seen for $q<1$. The dynamics then promotes minority opinions. This leads to long-lived states of opinion coexistence, stabilised by deterministic drift. These these states are described by the stable fixed points of the underlying rate equations.   

The extinction of opinions is driven by fluctuations. Individual realisations go through a sequence of extinctions, and the corresponding plateaux for the density of active interfaces. Unlike in the linear model, all realisations are seen to have the same quantitative sequence of plateaux. For all-to-all interaction, the interface density at these plateaux can be obtained from the fixed points of the rate equations. The escape times from the long-lived states are found to depend on $q$, and to scale exponentially in the population size $N$. 

 When $q>1$ the deterministic pull favours majority opinions, and drives the system to consensus. Only the consensus states are linearly stable fixed points of the rate equations. All other fixed points are either unstable or saddle points. For balanced initial conditions noise-driven symmetry breaking is required in the early phases of the dynamics. Other than this, noise plays virtually no role for the ordering, and there no further plateaux on the way to consensus. Broadly, the system follows the deterministic rate equations. 

Closed-form estimates for the consensus time can be found for the model with two opinion states, all-to-all interaction and when either $q=2$ and $q=3$. Approximations are also possible via an expansion when $q$ takes values just above one.  The time-to-consensus is found to scale logarithmically in $N$.

 \subsubsection{Noise-driven ordering: absence of single exponential decay of the average density for $q<1$}

  We have also studied the evolution of the average density of active interfaces across realizations. In the linear multi-state model, this quantity is known to show one single exponential decay law throughout the ordering process, with a time constant that is independent of $M$~\cite{msvm}.
  
 For $q>1$ we also find exponential decay at long times, but unlike in the linear model this is driven by the underlying deterministic pull rather than by noise.
 
 For $q<1$ the evolution of the average interface density is more intricate. As in the linear model, consensus is reached via a sequence of fluctuation-driven extinctions. In the nonlinear model the escape time from each of metastable states can be approximated as a Poisson process, giving rise to a sequence of exponential decay laws. Nonetheless, no overall single exponential decay, valid for the entire ordering process, emerges for $q<1$. This confirms the exceptional character of the linear model, where there is no deterministic drift. In the linear model the mean residence time in each plateau, and the mean reduction in interface density upon extinction of an opinion conspire to produce one single exponential decay law \cite{msvm}.
\subsubsection{Ordering dynamics on graphs and homogeneous pair approximation}
We have also compared the behaviour of nonlinear voter models on graphs with those with all-to-all interaction. We find broadly similar behavior, but the mean consensus time and other time scales of the ordering process can depend on the structure of the network. There are further quantitative difference, for example for the interface density at the different plateaux.

To describe nonlinear multi-state voter models on uncorrelated graphs we have developed a homogeneous pair approximation. Unlike previous approaches \cite{jed_qvm,min,min2,raducha}, our approximation does not rely on $q$ taking integer values. Instead, our approximation can be used for any positive real value of $q$, including in particular the range $0<q<1$. We also note that the only information some of the previous approximations use about the network is the mean degree~\cite{jed_qvm,min,min2,raducha}. Our approach instead relies on more detailed information on the graph, such as the mean inverse degree and/or higher-order moments of the degree distribution. 

We find that our approximation captures the ensuing plateaux well, and appears to be more accurate than previously existing pair approximations which implicitly rely on large mean degrees.

\subsection{Conclusions}
If anyone still doubts that simple dynamics such as those of the voter model can give rise to surprisingly complex and diverse behaviour, then our study is one more piece of evidence to the contrary (along with a large body of previous work). We have systematically studied the effects of nonlinearity in voter models with multiple opinion states, and shown that several distinct types of ordering emerge, depending on the exact type of the nonlinearity. In particular, deterministic drift can either promote order and consensus (when $q>1$), or co-existence (when $q<1$). The former leads to a quick ordering process, and the latter to long-lived metastable states. Our analysis also highlights the special character of the conventional linear voter model ($q=1$), where there is no deterministic pull away from or towards any opinion. The ordering is then purely driven by noise, and, unlike in the model with $q\neq 1$, different realizations undergo quite different paths towards order ih the linear model. Social scientists would perhaps refer to this as `historical contingency' \cite{blount}, i.e., the idea that the paths a system can take are often determined by randomness rather than a deterministic sequence of inevitable events.

Certainly, as we have shown, the statistical physics of the models with sub-linear, linear or supra-linear imitation dynamics are very different from one another. The aim of our work was to contrast these with one another, both in systems with all-to-all interaction and on graphs. 

Give that nonlinearity in imitation dynamics can significantly affect the formation of consensus in voter models, it is natural to ask what the consequences of nonlinear imitation might be on other models of social dynamics. Extensions of the multi-state voter model itself could include more general nonlinear rules by which an agent chooses an opinion from a sample of the states of several other agents. 

Some of these extensions may add realism to existing models of opinion dynamics. But even if this is not the main motivation, it seems likely that work along those lines will reveal new types of ordering, and therefore add to the statistical physics of interacting agent systems with absorbing states.

\acknowledgments{
Partial financial support has been received from the Agencia Estatal de Investigaci\'on (AEI, MCI, Spain) MCIN/AEI/10.13039/501100011033 and Fondo Europeo de Desarrollo Regional (FEDER, UE) under Projects APASOS (PID2021-122256NB-C21, PID2021-122256NB-C22) and the María de Maeztu Program for units of Excellence in R\&D, grant CEX2021-001164-M. FV thanks Prof. Juli\'an Fern\'andez Bonder for useful discussions and CONICET (Argentina) for continued support.}

\begin{appendix}

\section{Linear stability analysis for rate equations}
\subsection{Jacobian and full-coexistence fixed point}
The rate equations for the model with $M$ opinions are 
\BE
\dot x_\alpha & = & (1-x_\alpha)x_\alpha^q - x_\alpha \sum_{\beta\neq \alpha,\beta\neq M} x_\beta^q \nonumber \\
&& - x_\alpha \left(1-\sum_{\beta\neq M} x_\beta\right)^q \nonumber \\
& \equiv & f_\alpha,
\EE
for $\alpha=1,\dots,M-1$. We note again that there are only $M-1$ degrees of freedom in the model with $M$ opinions, due to $\sum_{\alpha=1}^M x_\alpha=1$. We have thus used $x_M=1-\sum_{\beta\neq M} x_\beta$.

We have
\BE
\frac{\partial f_\alpha}{\partial x_\alpha}&=&-x_\alpha^q + q x_\alpha^{q-1}(1-x_\alpha)-\sum_{\beta\neq\alpha, \beta\neq M} x_\alpha^q \nonumber \\
&&- \left(1-\sum_{\beta\neq M} x_\beta\right)^q +qx_\alpha\left(1-\sum_{\beta_\neq M} x_\beta\right)^{q-1}, \nonumber \\
\EE
and, for $\beta\neq \alpha$, 
\BE
\frac{\partial f_\alpha}{\partial x_\beta}&=&  -q x_\alpha x_\beta^{q-1}+qx_\alpha\left(1-\sum_{\beta\neq M}x_\beta\right)^{q-1}. \nonumber \\
\EE

At the full-coexistence fixed point we have $x_\alpha=1/M$ for $\alpha=1,\dots,M-1$.  

After some algebra we then find
\BE
\frac{\partial f_\alpha}{\partial x_\alpha}&=&(q-1)\left(\frac{1}{M}\right)^{q-1},\nonumber \\
\frac{\partial f_\alpha}{\partial x_\beta}&=&0,
\EE
where $\beta\neq\alpha$ in the second line.

This means that the Jacobian at the full-coexistence fixed point is diagonal, with diagonal elements $(q-1) (1/M)^{q-1}$. Thus, this fixed point is stable when $q<1$, and unstable when $q>1$.

\subsection{Partial co-existence fixed point}
The Jacobian becomes singular for $q<1$ when one or more of the opinions have gone extinct. We therefore assess the stability of the partial co-existence and consensus fixed points with a different method.

First, we look at the fixed point at which one opinion has gone extinct, and where all other opinions are equally represented in the population. We choose the extinct opinion to be opinion $M$.

The point $[1/(M-1),\dots, 1/(M-1),0]$ is the full co-existence fixed point in a model with $M-1$ opinions. So, within the space of opinions $1$ to $M-1$ only, this fixed point is stable for $q<1$, and unstable for $q>1$. 

We only need to study what happens if we introduce a small fraction of agents, $\epsilon$, holding opinion $M$. We assume that opinion $\alpha$ ($\alpha=1,\dots,M-1$) is then held by a fraction $x_\alpha=1-y_\alpha \epsilon$, where $\sum_{\alpha=1}^{M-1}y_\alpha=1$. We assume that $\epsilon\ll 1$.

We then have, from the rate equations,
\BE
\dot \epsilon &=& (1-\epsilon) \epsilon^q - \epsilon \sum_{\alpha \neq M} (1-y_\alpha \epsilon )^q, \nonumber \\
&=& \epsilon^q - \epsilon^{q+1}- \epsilon \sum_{\alpha \neq M} (1- y_\alpha \epsilon )^q.\label{eq:aux_stab}
\EE
For $q<1$, the right-hand side is dominated by the term $\epsilon^q$, and therefore remains positive. This means that the fixed point is unstable against the re-introduction of opinion $M$.

For $q>1$, the lowest power of $\epsilon$ on the RHS of Eq.~(\ref{eq:aux_stab}) comes from the last term. More precisely, and given that $(1-y_\alpha \epsilon )^q\approx 1-qy_\alpha\epsilon$, we have $\dot\epsilon\approx -(M-1)\epsilon$. Hence, the fixed point is stable against the re-introduction of opinion $M$.

Once one opinion has become extinct, the model reduces to a nonlinear VM with $M-1$ states. The partial co-existence fixed point with $M-1$ opinions in the model with $M$ opinions  becomes the full co-existence fixed point of the model with $M-1$ opinions. The procedure to assess the stability of fixed points can then be iterated.

\subsection{Consensus fixed point}
Let us now look at the stability of the fixed point describing consensus on opinion $M$, against the introduction of the other opinions. We assume $x_M = 1-\epsilon $ and $x_\alpha=y_\alpha \epsilon$ for $\alpha=1,\dots,M-1$, with $\sum_{\alpha=1}^{M-1}y_\alpha=1$, where again $\epsilon\ll 1$.

We have (from the rate equation for $\dot x_M$),
\BE
-\dot\epsilon &=&(1-\epsilon)^q \epsilon - (1-\epsilon) \sum_{\alpha=1}^M (y_\alpha \epsilon)^q \nonumber \\
&\approx &\epsilon (1-q\epsilon) - (1-\epsilon) \epsilon^q\sum_{\alpha=1}^M y_\alpha^q,\label{eq:stab_aux2}
\EE
where we draw attention to the minus sign on the left-hand side (resulting from $\dot x_M=-\dot\epsilon$).

For $q<1$, the right-hand side of Eq.~(\ref{eq:stab_aux2}) is dominated by the term 
$-\epsilon^q$, and will hence be negative overall. Thus, $\dot \epsilon>0$, and the fixed point is unstable. For $q>1$ on the other hand, we have $\dot \epsilon\approx -\epsilon$, and hence, the fixed point is stable.

 \section{Large-degree pair approximation (LDPA) for multistate nonlinear voter models} 
 \label{ap:minM}

We discuss here a pair approximation along the lines of Ref.~\cite{min,raducha,jed_qvm}. This existing work is for the two-state model in random regular networks of degree $\mu$. As discussed in the main text, this approximation turns out to be valid for ER networks of sufficiently large mean degree. We here generalise this to an arbitrary number $M$ of opinion states.

The starting point is the following equation for the time evolution of the average density of active interfaces 
\be\label{eq:drho_dt_appendix}
\frac{d\rho}{dt} = \sum^M_\alpha \sum^M_{\beta \neq \alpha}x_\alpha P(\alpha \rightarrow \beta)\Delta \rho (\alpha \rightarrow \beta)
\ee
Here, $x_\alpha$ is the probability of picking a node in opinion state $\alpha$. $P(\alpha \rightarrow \beta) $ is the probability that a subsequent change of opinion $\alpha \rightarrow \beta$ occurs. In our model, this probability is the $q$-th power of the fraction of neighbours of the focal agent in opinion state $\beta$. In the spirit of a pair approximation, this fraction is given by the fraction of active links of type $\alpha\beta$ among the links of an agent of type $\alpha$. This, in turn, can be written as $\rho_{\alpha\beta}/(2x_\alpha)$. The change in the density of active links is, as before, $\Delta \rho (\alpha \rightarrow \beta) = \mu - 2n_\beta - \sum^M_{\gamma \neq \alpha,\beta} n_\gamma$. 

In order to approximate $\Delta\rho$, and following \cite{min,raducha,jed_qvm}, one can now assume that $q$ is integer. Then, $P(\alpha \rightarrow \beta)$ can be thought as the probability of finding a set of $q$ randomly chosen neighbours all in state $\beta$. If neighbours are chosen without repetition, then  at least $q$ of the focal individual's neighbors must be in state $\beta$. The remaining $\mu-q$ neighbors can be in any state. Given that the focal agent is in state $\alpha$, and the probability for a neighbour to be in state $\gamma$ is $\rho_{\alpha\gamma}/(2x_\alpha)$  Hence, we arrive at the following approximation, 
\begin{eqnarray}
\Delta \rho (\alpha \rightarrow \beta) &\approx& \mu 
- 2[q+(\mu-q)\rho_{\alpha\beta}/2(x_\alpha)] \nonumber \\
&&- \sum_{\gamma \neq \alpha,\beta} (\mu-q)\rho_{\alpha\gamma}/(2x_\alpha) \nonumber. 
\end{eqnarray}

In the stationary state we set $d\rho/dt=0$ in Eq.~(\ref{eq:drho_dt_appendix}), and use the fact that all opinions are equally distributed ($x_\alpha = 1/M$), as well as  $\rho_{\alpha\beta}/(2x_\alpha) = \rho/(M-1)$, and the fact that all flips contribute equally. We then have
\begin{eqnarray}
0 &=& M(M-1)\frac{1}{M} \left( \frac{\rho}{M-1} \right)^q \nonumber \\
 \\ &\times& \Biggl\{ \mu- 2 \left[ q+ \frac{(\mu-q) \rho}{(M-1)} \right] - (M-2) \frac{(\mu-q) \rho}{(M-1)} \Biggr\}. \nonumber \\
\end{eqnarray}
From this we find
\be
\rho_{\mbox{\tiny LD}}^* = \frac{(M-1) (\mu-2q)}{M(\mu-q)},
\ee
which is Eq.~(\ref{eq:min_generalM}) of the main text.

\section{Linear stability analysis for the case $M=3$ and $q=2$ in CG}
\label{ap:approach-M3-CG}

For $M=3$ and $q=2$, the rate equations~(\ref{eq:rateM}) reduce to
\begin{subequations} 
\begin{align}
    \dot{x_1}&=(1-x_1)x_1^2-x_1 \left[ x_2^2+(1-x_1-x_2)^2 \right], \\
    \dot{x_2}&=(1-x_2)x_2^2-x_2 \left[ x_1^2+(1-x_1-x_2)^2 \right],
\end{align}
\label{eq:dxdt}
\end{subequations}
where we have used $x_3=1-x_1-x_2$. 

We now study the linear stability of the fixed point at $x_1=1, x_2=0$. We write $x_1=1-\epsilon_1$ and $x_2=\epsilon_2$, and linearise in $\epsilon_1$ and $\epsilon_2$. We find
\be
    \frac{d \vec{\epsilon}}{dt} = A \vec{\epsilon},
\ee
where
\be
A \equiv 
\begin{pmatrix}
-1 & 0 \\
 0 & -1 
\end{pmatrix}
\ee
and $\vec{\epsilon}=(\epsilon_1,\epsilon_2)^T$.  Thus, $\vec{x}^*$ is linearly stable under a small perturbation in any direction.  Then, the time evolution of $x_1$ and $x_2$ after a perturbation from the fixed point is given by 
\begin{subequations} 
\begin{align}
    x_1(t) &\approx 1-[1-x_1(0)]e^{-t}, \\
    x_2(t) &\approx x_2(0) e^{-t}.
\end{align}
\end{subequations} 
From these equations we obtain that the density of active links $\rho=2(x_1x_2+x_1x_3+x_2x_3)=2[x_1x_2+(x_1+x_2)(1-x_1-x_2)]$ decays as 
\begin{eqnarray}
    \rho(t) \approx  2[1-x_1(0)]e^{-t} 
\end{eqnarray}
in the long time limit $t \gg 1$, after a small perturbation.  This exponential relaxation of $\rho$ is quoted in Sec.~\ref{sec:q1}.

\section{Approach to stable attractors for the case $M=2$ and $q=2$ on complex networks}
\label{ap:approach}

 Here we show that, for an arbitrary initial condition $m(0)>0$ and $\rho(0)>0$, $m(t) \to 1$ and $\rho(t) \approx \alpha [1-m(t)]$ asymptotically, that is, in the limit $t\gg 1$.

To show this, we first note that 
$dm/dt$ as given by the RHS of Eq.~(\ref{eq:dmdtq2}) does not have any zeros in $0<m<1$, is continuous in $m$ and in particular it is strictly positive ($c_1>0$ due to $\mu_{-1}<1$) for $0<m<1$ and $\rho \ne 0$. It is reasonable to conclude that $m$ tends to $1$ in the long-time limit.  

Then, using this result we find from Eq.~(\ref{eq:drdtq2}) that $d\rho/dt$ is negative for large $t$, given that the last term in Eq.~(\ref{eq:drdtq2}) dominates as $m$ approaches $1$.  We note that $c_4>0$ for the networks in our simulations (we have $\mu + 2\mu_{-1} >3$ for DRRGs and ER networks with mean degree $\mu>2$). Thus $\rho$ tends to $0$. Therefore, the point $(1,0)$ is an attractor of the dynamics for an arbitrary initial condition $m(0)>0$ and $\rho(0)>0$.  Note that $(1,0)$ and $(-1,0)$ are not fixed points because the RHS of Eqs.~(\ref{eq:q2m2}) are not defined for $m=\pm 1$ and $\rho=0$. 

Let us now define a new variable $u \equiv \rho/(1-m^2)$. We then have
\be
    \dot{u} = \frac{(1-m^2) \dot{\rho} + 2 m \rho \dot{m}}{(1-m^2)^2}, 
\ee
 
which becomes 
\be
    \dot{u} = u \Big\{ c_2 + c_3 u + \left[ (2c_1-c_4) m^2 - c_4\right] u^2 \Big\}
    \label{eq:dotu}
\ee

after replacing the expressions for $\dot{m}$ and $\dot{\rho}$ from Eqs.~(\ref{eq:q2m2}).  The fixed points of Eq.~(\ref{eq:dotu}) satisfy the relation
\be
    u \Big\{ \left[ (c_4-2c_1) m^2 + c_4\right] u^2 - c_3 u - c_2 \Big\} = 0.
\ee
The fixed point $u^*=0$ corresponds to the solutions $\rho=0$ and $m \ne \pm 1$. This represents the line of unstable fixed points $(m,0)$ (with $-1<m<1$) of Eqs.~(\ref{eq:q2m2}). 

The other two fixed points are 
\be
    u_{\pm} = \frac{c_3 \pm \sqrt{c_3^2+4 \left[ (c_4-2c_1) m^2 + c_4\right] c_2}}{2\left[ (c_4-2c_1) m^2 + c_4\right]}.
\ee
For $\mu>2$ we have $c_4>0$, and $c_2>0$, and thus $u_- \le 0$ for $-1 \le m \le 1$, independent of the sign of $(c_4-2c_1) m^2 + c_4$. However, the solution $u_-<0$ is non-physical because it gives $\rho<0$ for $0<m<1$, and $m>1$ for $\rho>0$.

Then, the physical solution $u_+>0$ is obtained under the condition $c_4>c_1$, which we find to be fulfilled for $\mu>4$ in DRRGs, and $\mu \gtrsim 3.2$ for ER networks. If these conditions hold, then we find in the $t \to \infty$ limit that $m \to 1$, and $\rho(t) \to \alpha [1-m(t)]$ to first order in $1-m \ll 1$. The constant $\alpha$ is given by $\alpha=2u_+(m=1)$, i.e., we have 
\be
    \alpha = \frac{c_3 +\sqrt{c_3^2+8 (c_4-c_1)c_2}}{2 (c_4-c_1)},
\ee
as quoted in Eq.~(\ref{alpha}) of the main text.  Finally, as we are interested in the behavior close to $(1,0)$, we can define a function $\epsilon(t)$ that satisfies $0< \epsilon(t) \ll 1$ and rewrite $m$ and $\rho$ to first order in $\epsilon$ as $m(t)=1-\epsilon(t)$ and $\rho(t) = \alpha [1-m(t)] = \alpha \epsilon(t)$.

\end{appendix}

\end{document}